\RequirePackage{fix-cm} 
\documentclass[a4paper, twoside, reqno, dvips, 12pt]{amsart}
\usepackage{fixltx2e}   

\usepackage{etex}

\usepackage[latin1]{inputenc}
\usepackage[T1]{fontenc}

\usepackage{esint}
\usepackage{dsfont}
\usepackage{xspace}
\usepackage{amsgen}
\usepackage{amsthm}
\usepackage{amssymb}
\usepackage{amsmath}
\usepackage{wasysym}
\usepackage{upgreek}
\usepackage{amsfonts}
\usepackage{stmaryrd}
\usepackage{mathtools}

\usepackage{relsize}
\usepackage[mathcal, mathscr]{euscript}

\usepackage{mathrsfs}
\DeclareMathAlphabet{\mathscrbf}{OMS}{mdugm}{b}{n}

\usepackage{a4wide}

\usepackage[dvipsnames, table]{xcolor}
\definecolor{bckg}{RGB}{20.8, 20.8, 20.8}
\definecolor{oneblue}{rgb}{0.0, 0.0, 0.85}
\definecolor{Lightblue}{RGB}{214, 214, 214}
\definecolor{bluepigment}{rgb}{0.2, 0.2, 0.6}
\definecolor{charcoal}{rgb}{0.21, 0.27, 0.31}
\definecolor{denimblue}{rgb}{0.08, 0.38, 0.74}
\definecolor{Lightgray}{rgb}{0.89, 0.89, 0.89}
\definecolor{darkgrey}{rgb}{0.273, 0.281, 0.30}
\definecolor{darkelectricblue}{rgb}{0.33, 0.41, 0.47}

\usepackage[sort&compress, comma, square, numbers]{natbib}

\usepackage{psfrag}
\usepackage{graphicx}
\usepackage{subfigure}
\usepackage{morefloats}
\usepackage{indentfirst}

\usepackage{acronym}
\usepackage{microtype}
\usepackage[labelsep=period,%
            labelfont={bf,sf,color=bluepigment},%
            justification=raggedright]{caption}

\usepackage[perpage, symbol]{footmisc}

\usepackage[usenames, dvipsnames, pdf]{pstricks}
\usepackage{epsfig}
\usepackage{pst-grad} 
\usepackage{pst-plot} 

\usepackage[colorlinks,
           urlcolor=oneblue,
           linkcolor=denimblue,
           citecolor=NavyBlue,
           bookmarksopen=false,
           pdfpagemode=UseNone,
           pagebackref]{hyperref}

\usepackage[explicit]{titlesec}

\titleformat{\section}[block]
  {\color{NavyBlue}\Large\sffamily\bfseries}
  {}
  {0.0em}
  {\colorbox{bckg!5}{\strut\parbox{\dimexpr\linewidth-2\fboxsep\relax}{\thesection. #1}}}
  [\vspace*{0.33em}]

\titleformat{name=\section,numberless}[block]
  {\color{NavyBlue}\Large\sffamily\bfseries}
  {}
  {0.0em}
  {\colorbox{bckg!5}{\strut\parbox{\dimexpr\linewidth-2\fboxsep\relax}{#1}}}
  [\vspace*{0.33em}]

\titleformat{\subsection}
  {\color{NavyBlue}\large\sffamily\bfseries}
  {}
  {0.0em}
  {\colorbox{bckg!5}{\parbox{\dimexpr\linewidth-2\fboxsep\relax}{\thesubsection. #1}}}
  [\vspace*{0.33em}]

\titleformat{name=\subsection,numberless}
  {\color{NavyBlue}\Large\sffamily\bfseries}
  {}
  {0em}
  {\colorbox{bckg!5}{\parbox{\dimexpr\linewidth-2\fboxsep\relax}{#1}}}
  [\vspace*{0.33em}]

\titleformat{\subsubsection}
  {\color{bluepigment}\sffamily\normalsize\bfseries}
  {\thesubsubsection}
  {0.5em}
  {#1}
  [\vspace*{0.33em}]

\titleformat{\paragraph}[runin]
  {\color{bluepigment}\sffamily\small\bfseries}
  {}
  {0em}
  {#1}

\titlespacing{\section}{0.0em}{1.5em plus 2pt minus 2pt}%
{1.0em plus 2pt minus 2pt}[0em]
\titlespacing{\subsection}{0.5em}{1.5em plus 2pt minus 2pt}%
{1.0em}[0em]
\titlespacing{\subsubsection}{0.5em}{1.5em plus 2pt minus 2pt}%
{1.0em plus 2pt minus 2pt}[0em]

\usepackage{titletoc}

\contentsmargin{0.5em}
\newlength{\tocsep} 
\titlecontents{section}[\tocsep]
  {\addvspace{10pt}\bfseries\sffamily}
  {\contentslabel[\thecontentslabel]{\tocsep}}
  {}
  {\ \titlerule*[0.75pc]{.}\ \thecontentspage}
  []
\titlecontents{subsection}[\tocsep]
  {\addvspace{8pt}\sffamily}
  {\contentslabel[\thecontentslabel]{\tocsep}}
  {}
  {\ \titlerule*[0.5pc]{.}\ \thecontentspage}
  []
\titlecontents*{subsubsection}[\tocsep]
  {\addvspace{2pt}\footnotesize\sffamily}
  {}
  {}
  {\ \titlerule*[0.35pc]{.}\ \thecontentspage}
  [\\*]

\makeatletter
\def\@setauthors{%
  \begingroup
  \def\thanks{\protect\thanks@warning}%
  \trivlist
  \centering\footnotesize \@topsep30\p@\relax
  \advance\@topsep by -\baselineskip
  \item\relax
  \author@andify\authors
  \def\\{\protect\linebreak}%
  \textsc{\normalsize\textcolor{darkelectricblue}{\authors}}%
  \ifx\@empty\contribs
  \else
    ,\penalty-3 \space \@setcontribs
    \@closetoccontribs
  \fi
  \endtrivlist
  \endgroup
}
\def\@settitle{\begin{center}%
  \baselineskip14\p@\relax
    \bfseries
    \textsc{\Large\textcolor{charcoal}{\@title}}
  \end{center}%
}
\makeatother

\usepackage{enumitem}
\setlist[description]{%
  topsep=30pt,               
  itemsep=5pt,               
  font={\bfseries\sffamily\color{NavyBlue}}, 
}

\usepackage{fancyhdr}
\usepackage{lastpage}

\newcommand*\Title{\textcolor{bluepigment}{Dispersive shallow water wave modelling. Part III}}
\newcommand*\Authors{\textcolor{bluepigment}{G.~Khakimzyanov, D.~Dutykh \& Z.~Fedotova}}
\newcommand*{\plogo}{\textcolor{gray}{{\texttt{arXiv.org} / \textsc{hal}}}} 

\numberwithin{equation}{section}

\newtheorem{remark}{Remark}

\newcommand{\up}[1]{$^{\mathrm{\small\textsf{#1}}}$} 

\newcommand{\w}{\upvarpi}
\newcommand{\dr}{\dot{r}}
\newcommand{\G}{\Upsilon}
\newcommand{\ga}{\upgamma}
\newcommand{\om}{\upomega}
\newcommand{\rb}{\check{r}}
\newcommand{\rs}{\tilde{r}}
\newcommand{\pc}{\check{p}}
\newcommand{\si}{\varsigma}

\newcommand{\R}{\mathds{R}}
\newcommand{\T}{\mathbb{T}}

\renewcommand{\tau}{\uptau}
\newcommand{\ud}{\mathrm{d}}

\newcommand{\D}{\mathscr{D}}

\renewcommand{\phi}{\varphi}
\newcommand{\J}{\mathcal{J}}
\newcommand{\V}{\mathcal{V}}
\newcommand{\W}{\mathscr{W}}
\newcommand{\Gm}{\mathbb{G}}
\newcommand{\Dd}{\mathcal{D}}
\newcommand{\Uc}{\mathcal{U}}
\newcommand{\Gg}{\mathscr{G}}
\newcommand{\Pp}{\mathscr{P}}
\newcommand{\Rr}{\mathscr{R}}
\newcommand{\Ss}{\mathscr{S}}

\newcommand{\ups}{\upupsilon}
\renewcommand{\beta}{\upbeta}
\newcommand{\U}{\mathscrbf{U}}
\renewcommand{\O}{\mathcal{O}}
\newcommand{\ro}{\mathring{r}}
\newcommand{\hd}{\mathring{h}}
\renewcommand{\H}{\mathcal{H}}
\renewcommand{\leq}{\leqslant}

\renewcommand{\P}{\mathcal{P}}
\renewcommand{\S}{\mathcal{S}}
\newcommand{\Sr}{\mathfrak{S}}
\newcommand{\Maltese}{\bigstar}
\newcommand{\Vv}{\mathscrbf{V}}
\renewcommand{\alpha}{\upalpha}

\newcommand{\g}{\boldsymbol{g}}
\newcommand{\x}{\boldsymbol{x}}
\newcommand{\eps}{\upvarepsilon}
\newcommand{\Fo}{\boldsymbol{F}}
\renewcommand{\Pi}{\mathfrak{P}}
\renewcommand{\u}{\boldsymbol{u}}
\renewcommand{\v}{\boldsymbol{v}}
\renewcommand{\lambda}{\uplambda}
\newcommand{\dphi}{\dot{\varphi}}
\newcommand{\ip}{{\,i^{\,\prime}}}
\newcommand{\jp}{{\,j^{\,\prime}}}
\newcommand{\Op}{{\,0^{\,\prime}}}
\newcommand{\dtheta}{\dot{\theta}}
\newcommand{\const}{\mathrm{const}}
\newcommand{\etad}{\mathring{\eta}}
\newcommand{\Ud}{\boldsymbol{U}_d\,}
\newcommand{\Vd}{\boldsymbol{V}_d\,}
\newcommand{\St}{\grave{\mathcal{S}}}
\newcommand{\dlambda}{\dot{\uplambda}}
\newcommand{\ap}{{\,\alpha^{\,\prime}}}
\newcommand{\gp}{{\,\gamma^{\,\prime}}}
\newcommand{\ros}{{\ro}^{\,\mathrm{s}}}
\newcommand{\rob}{{\ro}_{\,\mathrm{b}}}
\newcommand{\e}{\boldsymbol{\mathrm{e}}}
\newcommand{\Srs}{\mathfrak{S}^{\,\star}}
\renewcommand{\i}{\boldsymbol{\mathrm{i}}}
\newcommand{\Uu}{\boldsymbol{\mathcal{U}}}
\newcommand{\Vu}{\boldsymbol{\mathcal{V}}}
\newcommand{\bomega}{\boldsymbol{\upomega}}

\newcommand{\Pnh}{\power}
\newcommand{\pb}{\varrho}
\newcommand{\A}{\mathscrbf{A}}
\newcommand{\B}{\mathscrbf{B}}
\newcommand{\Cs}{\mathscrbf{C}}
\newcommand{\vO}{\boldsymbol{0}}
\newcommand{\vbeta}{\boldsymbol{\beta}}
\newcommand{\valpha}{\boldsymbol{\alpha}}

\newcommand{\power}{\raisebox{.15\baselineskip}{\Large\ensuremath{\wp}}}

\newcommand{\ie}{\emph{i.e.}\/ }
\newcommand{\eg}{\emph{e.g.}\/ }
\newcommand{\etc}{\emph{etc.}\/}
\newcommand{\etal}{\emph{et al.}\/ }

\renewcommand{\div}{\grad\scal}

\newcommand{\otimesf}{\bar{\otimes}}
\newcommand{\divf}{\bar{\grad}\scal}
\newcommand{\scal}{\boldsymbol{\cdot}}
\newcommand{\grad}{\boldsymbol{\nabla}}
\newcommand{\rot}{\mathop{\mathrm{rot}}}

\newcommand{\abs}[1]{\lvert\, #1\, \rvert}
\newcommand{\sign}{\mathop{\mathrm{sign}}}

\newcommand{\pd}[2]{\frac{\partial\/ #1}{\partial\/ #2}}
\newcommand{\od}[2]{\frac{\mathrm{d}\/ #1}{\mathrm{d}\/#2}}

\newcommand{\eqdef}{\mathop{\stackrel{\,\mathrm{def}}{:=}\,}}
\newcommand{\defeq}{\mathop{\stackrel{\,\mathrm{def}}{=:}\,}}

\usepackage{acronym}

\begin{document}

\title[\Title]{Dispersive shallow water wave modelling. Part III: Model derivation on a globally spherical geometry}

\author[G.~Khakimzyanov]{Gayaz Khakimzyanov}
\address{\textbf{G.~Khakimzyanov:} Institute of Computational Technologies, Siberian Branch of the Russian Academy of Sciences, Novosibirsk 630090, Russia}
\email{Khak@ict.nsc.ru}

\author[D.~Dutykh]{Denys Dutykh$^*$}
\address{\textbf{D.~Dutykh:} LAMA, UMR 5127 CNRS, Universit\'e Savoie Mont Blanc, Campus Scientifique, 
73376 Le Bourget-du-Lac Cedex, France}
\email{Denys.Dutykh@univ-savoie.fr}
\urladdr{http://www.denys-dutykh.com/}
\thanks{$^*$ Corresponding author}

\author[Z.~I.~Fedotova]{Zinaida Fedotova}
\address{\textbf{Z.~I.~Fedotova:} Institute of Computational Technologies, Siberian Branch of the Russian Academy of Sciences, Novosibirsk 630090, Russia}
\email{zf@ict.nsc.ru}



\begin{titlepage}
\thispagestyle{empty} 
\noindent
{\Large Gayaz \textsc{Khakimzyanov}}\\
{\it\textcolor{gray}{Institute of Computational Technologies, Novosibirsk, Russia}}
\\[0.02\textheight]
{\Large Denys \textsc{Dutykh}}\\
{\it\textcolor{gray}{CNRS--LAMA, Universit\'e Savoie Mont Blanc, France}}
\\[0.02\textheight]
{\Large Zinaida \textsc{Fedotova}}\\
{\it\textcolor{gray}{Institute of Computational Technologies, Novosibirsk, Russia}}
\\[0.08\textheight]

\vspace*{1cm}

\colorbox{Lightblue}{
  \parbox[t]{1.0\textwidth}{
    \centering\huge\sc
    \vspace*{0.7cm}
    
    \textcolor{bluepigment}{Dispersive shallow water wave modelling. Part III: Model derivation on a globally spherical geometry}
    
    \vspace*{0.7cm}
  }
}

\vfill 

\raggedleft     
{\large \plogo} 
\end{titlepage}


\newpage
\thispagestyle{empty} 
\par\vspace*{\fill}   
\begin{flushright} 
{\textcolor{denimblue}{\textsc{Last modified:}} \today}
\end{flushright}


\newpage
\maketitle
\thispagestyle{empty}


\begin{abstract}

The present article is the third part of a series of papers devoted to the shallow water wave modelling. In this part we investigate the derivation of some long wave models on a deformed sphere. We propose first a suitable for our purposes formulation of the full \textsc{Euler} equations on a sphere. Then, by applying the depth-averaging procedure we derive first a new fully nonlinear weakly dispersive base model. After this step we show how to obtain some weakly nonlinear models on the sphere in the so-called \textsc{Boussinesq} regime. We have to say that the proposed base model contains an additional velocity variable which has to be specified by a closure relation. Physically, it represents a dispersive correction to the velocity vector. So, the main outcome of our article should be rather considered as a whole family of long wave models.


\bigskip\bigskip
\noindent \textbf{\keywordsname:} motion on a sphere; long wave approximation; nonlinear dispersive waves; spherical geometry; flow on sphere \\


\end{abstract}


\newpage
\thispagestyle{empty}
\tableofcontents
\thispagestyle{empty}


\newpage
\section{Introduction}

Recent mega-tsunami events in Sumatra 2004 \cite{Lay, Ammon2005, Syno2006} and in Tohoku, Japan 2011 \cite{Mori2011, Grilli2012} required the simulation of tsunami wave propagation on the global trans-oceanic scale. Moreover, similar catastrophic events in the future are to be expected in these regions \cite{McCloskey2008}. The potential tsunami hazard caused by various seismic scenarii can be estimated by extensive numerical simulations. During recent years the modelling challenges of tsunami waves have been extensively discussed \cite{Murty2006, DGK}. On such scales the effects of Earth's rotation and geometry might become important. Several authors arrived to this conclusion, see \eg \cite{Tkalich2007, Grilli2007}. There is an intermediate stage where the model is written on a tangent plane to the sphere in a well-chosen point. In the present study we consider the globally spherical geometry without such local simplifications.

The direct application of full hydrodynamic models such as \textsc{Euler} or \textsc{Navier}--\textsc{Stokes} equations does not seem realistic nowadays. Consequently, approximate mathematical models for free surface hydrodynamics on rotating spherical geometries have to be proposed. This is the main goal of the present study. The existing (dispersive and non-dispersive) shallow water wave models on a sphere will be reviewed below. Nowadays, hydrostatic models are mostly used on a sphere \cite{Zeitlin2007, Tort2014}. The importance of frequency dispersion effects was underlined in \eg \cite{Tappin2008}. Their importance has been realized for tsunami waves generated by sliding/falling masses \cite{Ward, Beisel2012, Dutykh2012, Dutykh2011d}. However, we believe that on global trans-oceanic scales frequency dispersion effects might have enough time to accumulate and, hence, to play a certain r\^ole. Finally, the topic of numerical simulation of these equations on a sphere is another important practical issue. It will be addressed in some detail in the following (and the last) Part~{IV} \cite{Khakimzyanov2016b} of the present series of papers entirely devoted to shallow water wave modelling.

Shallow water equations describing long wave dynamics on a (rotating) sphere have been routinely used in the fields of Meteorology and Climatology \cite{Zeitlin2007}. Indeed, there exist many similarities in the construction of approximate models of atmosphere and ocean dynamics \cite{Marchuk1974}. The derivation of these equations by depth-averaging can be found in the classical monograph \cite{Haltiner1980}. The main numerical difficulties here consist mainly in (structured) mesh generation on a sphere and treating the degeneration of governing equations at poles (the so-called poles problem). So far, the finite differences \cite{Lanser1999, Liska2001} and spectral methods \cite{Boyd2000} were the most successful in the numerical solution of these equations. Our approach to these problems will be described in \cite{Khakimzyanov2016b}.

It is difficult to say who was the first to apply Nonlinear Shallow Water Equations (NSWE) on a sphere to the problems of Hydrodynamics. Contrary to the Meteorology, where the scales are planetary from the outset and the spherical coordinates are introduced even on local scales \cite{Lynch2014}, in surface wave dynamics people historically tended to use local \textsc{Cartesian} coordinates. However, the need to simulate trans-oceanic tsunami wave propagation obliges us to consider spherical and Earth's rotation effects. We would like to mention that in numerical modelling of water waves on the planetary scale the problem of poles does not arise since these regions are covered with the ice. Thus, the flow cannot take place there.

In \cite{Williamson1992} one can find various forms of shallow water equations on a sphere along with standard test cases to validate numerical algorithms. The standard form of Nonlinear Shallow Water Equations (NSWE) in the spherical coordinates $O\,\lambda\,\phi\,r$ is
\begin{align*}
  \H_{\,t}\ +\ \div[\,\H\,\u\,]\ &=\ 0\,, \\
  (\H\,u)_{\,t}\ +\ \div[\,\H\,u\,\u\,]\ &=\ \Bigl(\digamma\ +\ \frac{u}{R}\;\tan\phi\Bigr)\;\H\,v\ -\ \frac{g\,\H}{R\cos\phi}\;\pd{\eta}{\lambda}\,, \\
  (\H\,v)_{\,t}\ +\ \div[\,\H\,v\,\u\,]\ &=\ -\Bigl(\digamma\ +\ \frac{u}{R}\;\tan\phi\Bigr)\;\H\,u\ -\ \frac{g\,\H}{R}\;\pd{\eta}{\phi}\,. \\
\end{align*}
Here $\H(\lambda,\,\phi,\,t)\ \eqdef\ \bigl(\eta\ +\ d\bigr)\,(\lambda,\,\phi,\,t)$ is the total water depth and $\u(\lambda,\,\phi,\,t)$ is the linear speed vector with components
\begin{equation*}
  \u\,(\lambda,\,\phi,\,t)\ \eqdef\ \Bigl(R\cos(\phi)\cdot\dlambda,\, R\,\dphi\Bigr)\,,
\end{equation*}
where the over dot denotes the usual derivative with respect to time, \ie $\dot{(\cdot)}\ \eqdef\ \od{(\cdot)}{t}\,$. Function $d$ specifies the bottom bathymetry shape and $\digamma\ \eqdef\ 2\,\Omega\,\sin\phi$ is \textsc{Coriolis}'s parameter, $\Omega$ being the Earth constant angular velocity. The constant $g$ is the absolute value of usual gravity acceleration. The divergence operator in spherical coordinates is computed as
\begin{equation*}
  \div\bigl((\cdot)_1,\,(\cdot)_2\bigr)\ \eqdef\ \frac{1}{R\,\cos\phi}\;\Bigl[\,\pd{(\cdot)_1}{\lambda}\ +\ \pd{\bigl(\cos\phi\,(\cdot)_2\bigr)}{\phi}\,\Bigr]\,.
\end{equation*}
The right hand sides of the last two NSW equations contain the \textsc{Coriolis} effect, additional terms due to rotating coordinate system and hydrostatic pressure gradient. Recently a new set of NSW equations on a sphere was derived \cite{Cherevko2009a, Cherevko2009} including also the centrifugal force due to the Earth rotation. The applicability range of this model was discussed \cite{Cherevko2009a} and some stationary solutions are provided \cite{Cherevko2009}. NSWE on a sphere are reported in \cite{Murty2006} in a non-conservative form and including the bottom friction effects. The derivation of these equations can be found in \cite{Kowalik1993, Kolar1994}. In earlier attempts such as \cite{Murty1984} NSWE did not include terms
\begin{equation*}
  \Bigl(\frac{u}{R}\;\tan\phi\Bigr)\;\H\,v\ \qquad\ \mbox{ and }\ \qquad\ \Bigl(\frac{u}{R}\;\tan\phi\Bigr)\;\H\,u\,.
\end{equation*}
We remark however that the contribution of these terms might be negligible for tsunami propagation problems. This system of NSWE is implemented, for example, in the code \texttt{MOST}\footnote{Method Of Splitting Tsunami (MOST)} \cite{Titov1997}. The need to include dispersive effects was mentioned in several works. In \cite{Tkalich2007} linear dispersive terms were added to NSWE and this model was integrated in \texttt{TUNAMI-N2} code. This numerical model allowed the authors to model the celebrated \textsc{Sumatra} 2004 event \cite{Syno2006}. In another work published the same year \textsc{Grilli} \etal (2007) \cite{Grilli2007} outlined the importance to work in spherical coordinates even if in \cite{Grilli2007} they used the \textsc{Cartesian} version of the code \texttt{FUNWAVE}. This goal was achieved six years later and published in \cite{Kirby2013}. \textsc{Kirby} \etal used scaling arguments to introduce two small parameters ($\si$ and $\mu$ in notation of our study). A weakly nonlinear and weakly dispersive \textsc{Boussinesq}-type model was given and effectively used in \cite{Lovholt2008, Lovholt2010}. However, the authors did not publish the derivation of these equations. Moreover, they included a free parameter which can be used to improve the dispersion relation properties, even if this modification may appear to be rather ad-hoc without a proper derivation to justify it.

The systematic derivation of fully nonlinear models on a sphere was initiated in our previous works \cite{Fedotova2010, Fedotova2011a, Fedotova2014c, Shokin2015}. In this work we would like to combine and generalize the existing knowledge on the derivation of dispersive long wave models in the spherical geometry including rotation effects. We cover the fully and weakly nonlinear cases. The relation of our developments to existing models is outlined whenever it is possible. The derivation in the present generality has not been reported in the literature before.

The present study is organized as follows. In Section~\ref{sec:Euler} we present the full \textsc{Euler} equations on an arbitrary moving coordinate system. The modified scaled \textsc{Euler} equations are given in Section~\ref{sec:mEuler}. The base nonlinear dispersive wave model is then derived in Section~\ref{sec:nld} from the modified \textsc{Euler} equations. The base model has to be provided with a closure relation. Two particular and popular choices are given in Sections~\ref{sec:depth} and \ref{sec:surf}. Finally, the main conclusions and perspectives of the present study are outlined in Section~\ref{sec:disc}. As a reminder, in Appendix~\ref{app:tensor} we explain the notations and provide all necessary information from tensor analysis used in our study.


\section{Euler equations}
\label{sec:Euler}

\begin{figure}
  \centering
  \includegraphics[width=0.79\textwidth]{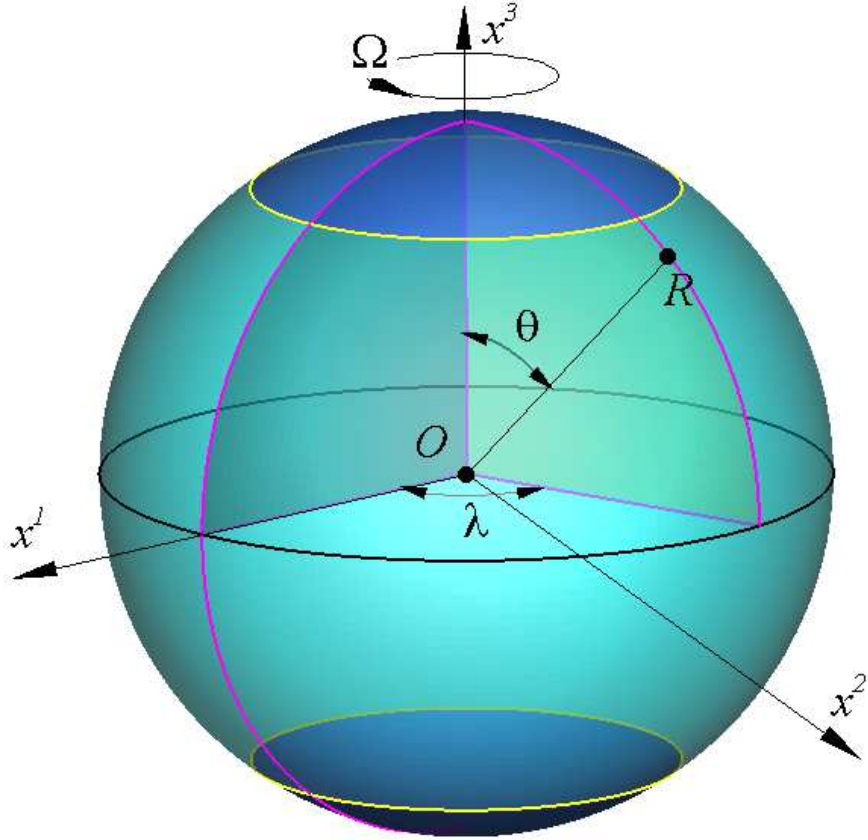}
  \caption{\small\em Cartesian and spherical coordinates used in this study.}
  \label{fig:sketch}
\end{figure}

The full \textsc{Euler} equations in spherical coordinates can be found in many works (see \eg the classical book \cite{Kochin1965}). However, for our purposes we prefer to have a more compact form of these equations. It will be derived in the present Section departing from \textsc{Euler} equations written in a standard \textsc{Cartesian} coordinate system $O\, x^1\, x^2\, x^3$. We assume that the axis $O\,x^3$ coincides with the rotation axis and points vertically upwards to the North pole. In this setting the coordinate plane $O\, x^1\, x^2$ coincides with the celestial equator. The definition of the employed \textsc{Cartesian} and spherical (curvilinear) coordinate systems is illustrated in Figure~\ref{fig:sketch}.

Moreover, we introduce a virtual sphere of radius $R$ whose center coincides with the center of Earth, $R$ being the mean Earth's radius. This sphere rotates with the angular velocity $\Omega$. We shall need this object below for the derivation of the base shallow water model. The real planet shape does not have to be spherical. We only assume that its geometry is globally spherical and can be obtained as a continuous deformation of the virtual sphere (shown in blue in Figure~\ref{fig:sketch}).

Among all volumetric forces we consider only the Newtonian gravity $\g$ directed towards the center of the rotating sphere. In other words, the force acting on a fluid particle located at the point $\x\ =\ \bigl(x^1,\,x^2,\,x^3\bigr)$ has the following expression:
\begin{equation*}
  \g\ =\ -g\;\frac{\x}{\abs{\x}}\ =\ -\frac{g}{\abs{\x}}\;\Bigl(x^1\,\i_1\ +\ x^2\,\i_2\ +\ x^3\,\i_3\Bigr)\,,
\end{equation*}
where $\bigl\{\i_{\,\alpha}\bigr\}_{\alpha\,=\,1}^{3}$ are unitary vectors of the \textsc{Cartesian} coordinate system. In the derivation of the base model we shall assume that the liquid layer depth is much smaller than Earth's (mean) radius $R\,$. However, for the \textsc{Euler} equations this assumption is not really needed. Moreover, we assume that the liquid is homogeneous, thus liquid density $\rho\ \equiv\ \const\,$. Moreover, for the sake of simplicity we assume that the gravity acceleration $g\ =\ \abs{\g}$ is also constant throughout the fluid bulk\footnote{The authors are not aware of any study in the field of Hydrodynamics where this assumption was not adopted.}. Under these conditions the equations which describe the motion of an ideal incompressible fluid are well known:
\begin{align}\label{eq:EulerCart1}
  \pd{\bigl(\rho\,U_{\,\alpha}\bigr)}{x^{\,\alpha}}\ &=\ 0\,, \\
  \pd{\bigl(\rho\,U_{\,\beta}\bigr)}{t}\ +\ \pd{\bigl(\rho\,U_{\,\beta}\,U_{\,\alpha}\bigr)}{x^{\,\alpha}}\ +\ \pd{\P}{x^{\,\beta}}\ &=\ -\rho\,g\;\frac{x^{\,\beta}}{\abs{\x}}\;, \qquad \beta\ =\ 1,\,2,\,3\,. \label{eq:EulerCart2}
\end{align}
In equations above and throughout this study we adopt the summation convention over repeating lower and upper indices. Functions $\bigl\{U_{\,\alpha}\bigr\}_{\alpha\,=\,1}^{3}$ are \textsc{Cartesian} components of the fluid particles velocity vector and $\P$ is the fluid pressure.


\subsection{Euler equations in arbitrary moving frames of reference}

A curvilinear coordinate system $(q^0,\,q^1,\,q^2,\,q^3)$ is given by a regular bijective homomorphism (or even a diffeomorphism) onto a certain domain with \textsc{Cartesian} coordinates $(x^0,\,x^1,\,x^2,\,x^3)$. In the present study for the sake of convenience we give a different treatment to time and space coordinates, \ie
\begin{equation}\label{eq:moving}
  x^0\ =\ q^0\ =\ t\,, \qquad
  x^\alpha\ =\ x^\alpha\;\bigl(q^0,\,q^1,\,q^2,\,q^3\bigr)\,, \qquad
  \alpha\ =\ 1,\,2,\,3\,.
\end{equation}
More precisely, we assume that the mapping above satisfies the following conditions \cite{Milnor1997}:
\begin{itemize}
  \item The map is bijective
  \item The map \emph{and} its inverse are at least continuous (or even smooth)
  \item The Jacobian of this map is non-vanishing
\end{itemize}
More information on curvilinear coordinate systems is given in Appendix~\ref{app:tensor}. As one can see, the time variable $t$ is chosen to be the same in both coordinate systems. It does not have to change (at least in the Classical Mechanics). Consequently, a point $P$ having \textsc{Cartesian} (spatial) coordinates $(x^1,\,x^2,\,x^3)$ will have curvilinear coordinates $(q^1,\,q^2,\,q^3)\,$.

\begin{remark}
Before reading the sequel of this article we strongly recommend to read first Appendix~\ref{app:tensor} where we provide all necessary information from tensor analysis and we explain the system of notations used below.
\end{remark}

The system of equations \eqref{eq:EulerCart1}, \eqref{eq:EulerCart2} can be recast in a compact tensorial form as follows \cite{Kovenya1981}:
\begin{equation*}
  \div\T\ =\ \rho\,\Fo\,,
\end{equation*}
where $\T\ =\ \bigl\{T^{\ip\jp}\bigr\}_{\ip\jp\,=\,0}^{\,3}$ is the $2-$tensor and $\Fo\ =\ \bigl\{F^{\ip}\bigr\}_{\ip\,=\,0}^{\,3}$ is the $1-$tensor. The last form has the advantage of being coordinate frame invariant (\ie independent of coordinates provided that the components of tensors $\T$ and $\Fo$ are transformed according to some well-established rules). However, in the perspective of numerical discretization \cite{Khakimzyanov2016b}, one needs to introduce explicitly the coordinate system into notation to work with. It can be done starting from the components of tensors $\T$ and $\Fo$ in a \textsc{Cartesian} frame of reference:
\begin{equation*}
  T^{\Op\jp}\ =\ \rho\,U^{\Op}\,U^{\jp}\ \equiv\ \rho\,U^{\jp}\,, \qquad
  T^{\ap\jp}\ =\ \rho\,U^{\ap}\,U^{\jp}\ +\ \P\,\delta^{\ap\jp}\,,
\end{equation*}
\begin{equation*}
  F^{\Op}\ =\ 0\,, \qquad
  F^{\ap}\ =\ -g\;\frac{x^{\ap}}{\abs{\x}}\,.
\end{equation*}
We employ indices with primes in order to denote \textsc{Cartesian} components. For instance, $U^{\jp}$ is the $j$\up{th} component of the velocity vector in a \textsc{Cartesian} frame of reference and it can be computed as
\begin{equation*}
  U^{\jp}\ =\ \dot{x}^{\jp}\ \eqdef\ \od{x^{\jp}}{t}\,,
\end{equation*}
$U^{\Op}$ being equal to $1$ thanks to the choice \eqref{eq:moving}. In any other moving curvilinear frame of reference the components of tensor $\T$ can be computed according to formulas \eqref{eq:tensor}:
\begin{equation*}
  T^{\,0\,j}\ =\ \D_{\ip}^{\,0}\,\D_{\jp}^{\,j}\,T^{\ip\jp}\ =\ \D_{\Op}^{\,0}\,\D_{\jp}^{\,j}\,T^{\Op\jp}\ =\ \D_{\jp}^{\,j}\,T^{\Op\jp}\ =\ \D_{\jp}^{\,j}\,\bigl(\rho\,U^{\jp}\bigr)\ =\ \rho\,\V^{\,j}\,,
\end{equation*}
\begin{equation*}
  T^{\,\alpha\,0}\ =\ \D_{\ip}^{\,\alpha}\,\D_{\jp}^{\,0}\,T^{\ip\jp}\ =\ \D_{\ip}^{\,\alpha}\,\D^{\,0}_{\Op}\,T^{\ip\Op}\ =\ \D_{\ip}^{\,\alpha}\,\bigl(\rho\,U^{\ip}\bigr)\ =\ \rho\,\V^{\,\alpha}\,,
\end{equation*}
\begin{multline*}
  T^{\alpha\,\beta}\ =\ \D_{\ip}^{\,\alpha}\,\D_{\jp}^{\,\beta}\,T^{\ip\jp}\ =\ \D^{\,\alpha}_{\Op}\,\D_{\jp}^{\,\beta}\,T^{\Op\jp}\ +\ \D_{\ap}^{\,\alpha}\,\D_{\jp}^{\,\beta}\,T^{\ap\jp}\ =\\ \D_{\Op}^{\,\alpha}\,\bigl(\rho\,\V^{\,\beta}\bigr)\ +\ \D_{\ap}^{\,\alpha}\,\bigl(\rho\,U^{\ap}\,\V^{\,\beta}\bigr)\ +\ \P\cdot\D_{\ap}^{\,\alpha}\,\D_{\jp}^{\,\beta}\,\delta^{\ap\jp}\ =\\
  \rho\,\V^{\,\alpha}\,\V^{\,\beta}\ +\ \P\cdot\bigl(\D_{\ip}^{\,\alpha}\,\D_{\jp}^{\,\beta}\,\delta^{\ip\jp}\ -\ \D_{\Op}^{\,\alpha}\,\D_{\jp}^{\,\beta}\,\delta^{\Op\jp}\bigr)\ \equiv\\ \rho\,\V^{\,\alpha}\,\V^{\,\beta}\ +\ \P\cdot\bigl(g^{\,\alpha\,\beta}\ -\ g^{\,\alpha\,0}\,g^{\,\beta\,0}\bigr)\,,
\end{multline*}
where $\{g^{\,ij}\}_{i,\,j\,=\,0}^{3}$ are components of the contravariant metric tensor (defined in Appendix~\ref{app:tensor}) and $\V^{\,j}$ is the $j$\up{th} contravariant component of velocity in a curvilinear frame of reference:
\begin{equation}\label{eq:speed}
  \V^{\,j}\ =\ \dot{q}^{\,j}\ \eqdef\ \od{q^{\,j}}{t}\ =\ \D_{\jp}^{\,j}\,U^{\jp}\,.
\end{equation}
The components of tensor $\Fo$ are transformed as
\begin{equation*}
  F^{\,0}\ =\ \D_{\ip}^{\,0}\,F^{\ip}\ \equiv\ 0\,, \qquad
  F^{\,\alpha}\ =\ \D_{\ip}^{\,\alpha}\,F^{\ip}\ =\ -g\;\frac{x^{\ap}}{\abs{\x}}\;\D_{\ap}^{\,\alpha}\,.
\end{equation*}
We have obviously that $\V^{\,0}\ =\ 1\,$, $T^{\,0\,0}\ =\ \rho$ and in Appendix~\ref{app:tensor} we show that
\begin{equation*}
  g^{\,0\,\alpha}\ \equiv\ q_{\,t}^{\,\alpha}\,, \qquad
  g^{\,0\,\beta}\ \equiv\ q_{\,t}^{\,\beta}\,.
\end{equation*}
The expressions of $2-$tensor $\T$ elements along with the $1-$tensor $\Fo$ are used to write the full $\textsc{Euler}$ equations in an arbitrary curvilinear coordinate system. The following compact notation is already familiar to us:
\begin{equation*}
  (\div\T)^{\,i}\ =\ \rho\,F^{\,i}\,, \qquad i\ =\ 0,\,\ldots,\,3\,,
\end{equation*}
or using formula \eqref{eq:div} for the divergence operator we have
\begin{equation}\label{eq:jtij}
  \pd{\,(\J\,T^{\,i\,j})}{q^{\,j}}\ +\ \J\,\G_{\,jk}^{\,i}\,T^{\,j\,k}\ =\ \rho\,\J\,F^{\,i}\,, \qquad i\ =\ 0,\,\ldots,\,3\,.
\end{equation}
For instance, for $i\ =\ 0$ we obtain the mass conservation equation in an arbitrary frame of reference:
\begin{equation}\label{eq:cont}
  \pd{\,(\rho\,\J)}{t}\ +\ \pd{\,(\rho\,\J\,\V^{\,\alpha})}{q^{\,\alpha}}\ =\ 0\,.
\end{equation}
For $i\ =\ 1,\,2,\,3$ from equation \eqref{eq:jtij} one obtains the momentum conservation equations, which can be expanded by inserting expressions of $2-$tensor components $T^{\,i\,j}\,$:
\begin{multline}\label{eq:moment}
  \pd{\,(\rho\,\J\,\V^{\,\beta})}{t}\ +\ \pd{\,(\rho\,\J\,\V^{\,\alpha}\,\V^{\,\beta})}{q^{\,\alpha}}\ +\ \J\;\bigl(g^{\,\alpha\,\beta}\ -\ g^{\,\alpha\,0}\,g^{\,\beta\,0}\bigr)\;\pd{\P}{q^{\,\alpha}}\ +\ \rho\,\J\,\G_{\,j\,k}^{\,\beta}\,\V^{\,j}\,\V^{\,k}\ + \\ \P\cdot\underbrace{\Bigl[\,\J\;\pd{\,(g^{\,\alpha\,\beta} - g^{\,\alpha\,0}\,g^{\,\beta\,0})}{q^{\,\alpha}}\ +\ \bigl(g^{\,\alpha\,\beta} - g^{\,\alpha\,0}\,g^{\,\beta\,0}\bigr)\;\pd{\,\J}{q^{\,\alpha}}\ +\ \J\,\G_{\,\alpha\,\gamma}^{\,\beta}\;\bigl(g^{\,\alpha\,\gamma} - g^{\,\alpha\,0}\,g^{\,\gamma\,0}\bigr)\,\Bigr]}_{(\bigstar)}\ =\ \rho\,\J\,F^{\,\beta}\,,
\end{multline}
where $\beta\ =\ 1,\,2,\,3\,$. By using formula \eqref{eq:gij} to differentiate the components of the contravariant metric components $g^{\,\alpha\,\beta}$ along with formula \eqref{eq:jac} one can show that expression $(\bigstar)\ \equiv\ 0\,$. Consequently, the momentum conservation equations simplify substantially. However, this set of equations still represents an important drawback: each equation contains the derivative of the pressure with respect to all three coordinate directions $q^{\,\alpha}\,$, $\alpha\ =\ 1,\,2,\,3\,$. So, we continue to modify the governing equations. We take index $\nu\ \in\ \{1,\,2,\,3\}$ and we multiply the continuity equation \eqref{eq:cont} by the covariant metric tensor component $g_{\,0\,\nu}\,$. Then, momentum conservation equation \eqref{eq:moment} is multiplied by $g_{\,\beta\,\nu}$ and we sum up obtained expressions. As a result, we obtain the following equation:
\begin{multline*}
  \pd{\,(\rho\,g_{\,j\,\nu}\,\J\,\V^{\,j})}{t}\ +\ \pd{\,(\rho\,g_{\,j\,\nu}\,\J\,\V^{\,\alpha}\,\V^{\,j})}{q^{\,\alpha}}\ +\ \J\,g_{\,\beta\,\nu}\cdot\bigl(g^{\,\alpha\,\beta}\ -\ g^{\,\alpha\,0}\,g^{\,\beta\,0}\bigr)\;\pd{\P}{q^{\,\alpha}} \\-\ \rho\,\J\,\V^{\,j}\;\pd{g_{\,j\,\nu}}{t}\ -\ \rho\,\J\,\V^{\,\alpha}\,\V^{\,j}\;\pd{g_{\,j\,\nu}}{q^{\,\alpha}}\ +\ \rho\,g_{\,\beta\,\nu}\,\J\,\G_{\,j\,k}^{\,\beta}\,\V^{\,j}\,\V^{\,k} \ =\ \rho\,g_{\,j\,\nu}\,\J\,F^{\,j}\,.
\end{multline*}
Using relation \eqref{eq:inverse} we can show that
\begin{equation*}
  g_{\,\beta\,\nu}\cdot\bigl(g^{\,\alpha\,\beta}\ -\ g^{\,\alpha\,0}\,g^{\,\beta\,0}\bigr)\ \equiv\ \delta_{\,\nu}^{\,\alpha}\,.
\end{equation*}
Using the relations \eqref{eq:cocontra} between covariant and contravariant components of a vector one obtains also
\begin{equation}\label{eq:coco}
  g_{\,j\,\nu}\,F^{\,j}\ =\ F_{\,\nu}\,, \qquad
  g_{\,j\,\nu}\,\V^{\,j}\ =\ \V_{\,\nu}\,.
\end{equation}
Above, $F_{\,\nu}$ and $\V_{\,\nu}$ are covariant components of the force and velocity vectors (or $1-$tensors) respectively. Finally, we obtain the \emph{conservative} form the full \textsc{Euler} momentum equations in an arbitrary frame of reference:
\begin{equation*}
  \pd{\,(\rho\,\J\,\V_{\,\nu})}{t}\ +\ \pd{\,(\rho\,\J\,\V^{\,\alpha}\,\V_{\,\nu})}{q^{\,\alpha}}\ +\ \J\;\pd{\P}{q^{\,\nu}}\ +\ \rho\,\J\,\V^{\,j}\,\V^{\,k}\;\Bigl[\,g_{\,\beta\,\nu}\,\G_{\,j\,k}^{\,\beta}\ -\ \pd{\,g_{\,j\,\nu}}{q^{\,k}}\,\Bigr]\ =\ \rho\,\J\,F_{\,\nu}\,, \quad \nu\ =\ 1,\,2,\,3\,.
\end{equation*}
By using the continuity equation \eqref{eq:cont}, one can derive similarly the non-conservative form of the momentum equation:
\begin{equation}\label{eq:moment}
  \pd{\V_{\,\nu}}{t}\ +\ \V^{\,\alpha}\;\pd{\V_{\,\nu}}{q^{\,\alpha}}\ +\ \frac{1}{\rho}\;\pd{\P}{q^{\,\nu}}\ +\ \V^{\,j}\,\V^{\,k}\;\Bigl[\,\G_{\,j\,k,\;\nu}\ -\ \pd{g_{\,j\,\nu}}{q^{\,k}}\,\Bigr]\ =\ F_{\,\nu}\,,  \quad \nu\ =\ 1,\,2,\,3\,,
\end{equation}
where we used \textsc{Christoffel} symbols of the first kind for the sake of simplicity.


\subsection{Euler equations in spherical coordinates}

From now on we choose to work in spherical coordinates since the main applications of our work aim the Geophysical Fluid Dynamics on planetary scales. As we know the planets are not exactly spheres. Nevertheless, the introduction of spherical coordinates still simplifies a lot the analytical work.

Consider a spherical coordinate system $O\,\lambda\,\theta\,r$ with the origin placed in the center of a virtual sphere of radius $R$ rotating with constant angular speed $\Omega\,$. By $\lambda$ we denote the longitude whose zero value coincides with a chosen meridian. Angle $\theta$ is the colatitude defined as $\theta\ \eqdef\ \frac{\pi}{2}\ -\ \phi\,$, where $\phi$ is the geographical latitude. Finally, $r\ >\ 0$ is the radial coordinate. Since latitude $-\frac{\pi}{2}\ <\ \phi\ <\ \frac{\pi}{2}$, we have that $0\ <\ \theta\ <\ \pi\,$. However, we assume additionally that
\begin{equation*}
  \theta_0\ \leq\ \theta\ \leq\ \pi\ -\ \theta_0\,,
\end{equation*}
where $1\ \gg\ \theta_0\ =\ \const\ >\ 0$ is a small angle. In other words, we exclude the poles with their small neighbourhood\footnote{In Atmospheric sciences this assumption is not realistic, of course. However, in Hydrodynamics it is justified by natural ice covers around pole regions --- Arctic and Antarctic. So, water wave phenomena do not take place near Earth's poles.}. Spherical coordinates $q^{\,0}\ =\ t\,$, $q^{\,1}\ =\ \lambda\,$, $q^{\,2}\ =\ \theta\,$, $q^{\,3}\ =\ r\,$ and \textsc{Cartesian} coordinates $x^{\,0}\,$, $x^{\,1}\,$, $x^{\,2}\,$, $x^{\,3}\,$ are related by the following formulas:
\begin{align*}
  x^{\,0}\ &=\ t\,, \\
  x^{\,1}\ &=\ r\,\cos(\lambda\ +\ \Omega\,t)\,\sin\theta\,, \\
  x^{\,2}\ &=\ r\,\sin(\lambda\ +\ \Omega\,t)\,\sin\theta\,, \\
  x^{\,3}\ &=\ r\,\cos\theta\,.
\end{align*}
Using formula \eqref{eq:jacobian} it is not difficult to show that \textsc{Jacobian} of the transformation above is
\begin{equation}\label{eq:jacs}
  \J\ =\ -r^2\,\sin\theta\,.
\end{equation}
Similarly, using formulas \eqref{eq:1.14}, \eqref{eq:1.15}, one can compute covariant components of the metric tensor:
\begin{equation*}
  g_{\,0\,0}\ =\ 1\ +\ \Omega^2\,r^2\,\sin^2\theta\,, \quad
  g_{\,1\,0}\ \equiv\ g_{\,0\,1}\ =\ \Omega\,r^2\,\sin^2\theta\,, \quad
  g_{\,2\,0}\ =\ g_{\,0\,2}\ =\ g_{\,3\,0}\ =\ g_{\,0\,3}\ \equiv\ 0\,,
\end{equation*}
\begin{equation*}
  g_{\,1\,1}\ =\ r^2\,\sin^2\theta\,, \quad 
  g_{\,2\,2}\ =\ r^2\,, \quad
  g_{\,3\,3}\ =\ 1\,, \quad
  g_{\,\alpha\,\beta}\ =\ g_{\,\beta\,\alpha}\ \equiv\ 0\,,
\end{equation*}
with $\alpha,\,\beta\ =\ 1,\,2,\,3\,$ and $\alpha\ \neq\ \beta\,$. From formulas \eqref{eq:speed} we compute contravariant components of the velocity vector:
\begin{equation*}
  \V^{\,0}\ =\ 1\,, \quad
  \V^{\,1}\ =\ \dlambda\,, \quad
  \V^{\,2}\ =\ \dtheta\,, \quad
  \V^{\,3}\ =\ \dr\,.
\end{equation*}
The covariant components of the velocity vector $\{\V_{\,\alpha}\}_{\alpha\,=\,1}^{3}$ and the exterior volume force $\{F_{\,\alpha}\}_{\alpha\,=\,1}^{3}$ are computed thanks to relations \eqref{eq:coco}:
\begin{align*}
  \V_{\,1}\ &=\ g_{\,1\,0}\ +\ g_{\,1\,1}\,\V^{\,1}\ =\ \Omega\,r^2\,\sin^2\theta\ +\ r^2\,\sin^2\theta\;\dlambda\,, \\
  \V_{\,2}\ &=\ g_{\,2\,2}\,\V^{\,2}\ =\ r^2\,\dtheta\,, \\
  \V_{\,3}\ &=\ g_{\,3\,3}\,\V^{\,3}\ =\ \dr\,,
\end{align*}
and the force components are
\begin{equation*}
  F_{\,1}\ =\ F_{\,2}\ \equiv\ 0\,, \qquad
  F_{\,3}\ =\ -g\,.
\end{equation*}
Finally, by using the definition of \textsc{Christoffel} symbols of the first kind, we obtain the sequence of the following relations for the term
\begin{multline*}
  \V^{\,j}\,\V^{\,k}\;\Bigl[\,\G_{\,j\,k,\;\nu}\ -\ \pd{g_{\,j\,\nu}}{q^{\,k}}\,\Bigr]\ \equiv\\ \sum_{k\,=\,1}^{3}\sum_{j\,=\,0}^{k-1}\Bigl[\,2\,\G_{\,j\,k,\,\nu}\ -\ \pd{g_{\,j\,\nu}}{q^{\,k}}\ -\ \pd{g_{\,k\,\nu}}{q^{\,j}}\,\Bigr]\,\V^{\,j}\,\V^{\,k}\ +\ \sum_{j\,=\,0}^{3}\Bigl[\,\G_{\,j\,j,\,\nu}\ -\ \pd{g_{\,j\,\nu}}{q^{\,j}}\,\Bigr]\,(\V^{\,j})^{\,2}\ =\\ -\sum_{k\,=\,1}^{3}\sum_{j\,=\,0}^{k-1}\,\pd{g_{\,j\,k}}{q^{\,\nu}}\,\V^{\,j}\,\V^{\,k}\ -\ \frac{1}{2}\;\sum_{j\,=\,0}^{3}\,\pd{g_{\,j\,j}}{q^{\,\nu}}\,(\V^{\,j})^{\,2}\,.
\end{multline*}
Using the fact that the \textsc{Jacobian} $\J$ is time-independent (see formula \eqref{eq:jacs}), we obtain the full \textsc{Euler} equations governing the flow of a homogeneous incompressible fluid in spherical coordinates:
\begin{align}\label{eq:seuler1}
  \pd{(\rho\,\J\,\V^{\,\alpha})}{q^{\,\alpha}}\ &=\ 0\,, \\
  \pd{\V_{\beta}}{t}\ +\ \V^{\,\alpha}\;\pd{\V_{\,\beta}}{q^{\,\alpha}}\ +\ \frac{1}{\rho}\cdot\pd{\P}{q^{\,\beta}}\ &=\ \S_{\,\beta}\,, \qquad \beta\ =\ 1,\,2,\,3\,, \label{eq:seuler2}
\end{align}
where
\begin{equation*}
  \S_{\,\beta}\ =\ \begin{dcases}
    \ 0\,, & \beta\ =\ 1\,, \\
    \ \frac{(\Omega + \V^{\,1})^2}{2}\cdot\pd{g_{\,1\,1}}{\theta}\,, & \beta\ =\ 2\,, \\
    \ -g\ +\ \frac{(\Omega + \V^{\,1})^2}{2}\cdot\pd{g_{\,1\,1}}{r}\ +\ \frac{(\V^{\,2})^2}{2}\cdot\pd{g_{\,2\,2}}{r}\,, & \beta\ =\ 3\,.
  \end{dcases}
\end{equation*}
The derivatives of covariant components of the metric tensor can be explicitly computed to give:
\begin{align*}
  \pd{g_{\,1\,1}}{\theta}\ &=\ 2\,r^2\,\sin\theta\,\cos\theta\,, \\
  \pd{g_{\,1\,1}}{r}\ &=\ 2\,r\,\sin^2\theta\,, \\
  \pd{g_{\,2\,2}}{r}\ &=\ 2\,r\,.
\end{align*}
We notice that components $\S_{\,2,\,3}$ contain correspondingly the terms $r^2\,\Omega^{\,2}\,\sin\theta\,\cos\theta$ and $r\,\Omega^{\,2}\,\sin^2\theta\,$. They are due to the centrifugal force coming from the Earth rotation. The presence of this force causes, for instance, the deviation of the pressure gradient from the radial direction even in the quiescent fluid layer. This effect will be examined in the following Section.

\subsubsection{Equilibrium free surface shape}

When we worked on a globally flat space \cite{Khakimzyanov2016c}, the free surface elevation $y\ =\ \eta(\x,\,t)$ was measured as the excursion of fluid particles from the coordinate plane $y\ =\ 0\,$. This plane is chosen to coincide with the free surface profile of a quiescent fluid at rest. On a rotating sphere, the situation is more complex since the equilibrium free surface shape does not coincide, in general, with any virtual sphere of a radius $R\,$. It is the centrifugal force which causes the divergence from the perfectly symmetric spherical profile. So, when we work on a sphere, the free surface elevation will be also measured as the deviation from the equilibrium shape. In this Section we shall determine the equilibrium free surface profile $r\ =\ R\ +\ \eta_{\,0\,0}(\,\lambda,\,\theta)$ by using two natural conditions:
\begin{itemize}
  \item The equilibrium profile along with bottom are steady
  \item The pressure $\P$ on the free surface is constant. Since the flow is incompressible, this constant can be set to zero without any loss of generality.
\end{itemize}

\begin{figure}
  \centering
  \includegraphics[width=0.72\textwidth]{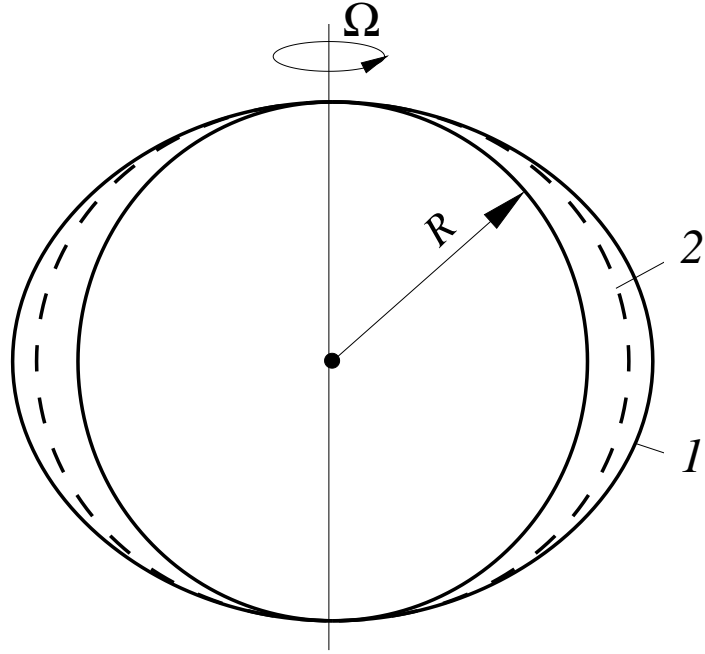}
  \vspace*{-3em}
  \caption{\small\em Stationary free surface profiles of a liquid layer over a virtual sphere of radius $R$ and rotating with constant angular speed $\Omega$: (1) solution of \textsc{Euler} equations; (2) solution of modified \textsc{Euler} equations.}
  \label{fig:sphere}
\end{figure}

In the case of a quiescent fluid (\ie all $\V^{\,\alpha}\ \equiv\ 0$), the full \textsc{Euler} equations of motion simply become
\begin{align*}
  \pd{\P}{\lambda}\ &=\ 0\,, \\
  \pd{\P}{\theta}\ &=\ \rho\,\Omega^2\,r^2\,\sin\theta\,\cos\theta\,, \\
  \pd{\P}{r}\ &=\ -\rho\,g\ +\ \rho\,\Omega^2\,r\,\sin^2\theta\,.
\end{align*}
The solution of these equations can be trivially obtained by successive integrations in each of spherical independent variables:
\begin{equation*}
  \P\ =\ -\rho\,g\,r\ +\ \frac{\rho\,r^2\,\Omega^{\,2}}{2}\;\sin^2\theta\ +\ C\,,
\end{equation*}
where $C\ =\ \const\ \in\ \R$ is an arbitrary integration constant which is to be specified later. Now we can enforce the dynamic boundary condition on the free surface, which states that the pressure $\P\ =\ 0$ vanishes at the free surface $r\ =\ R\ +\ \eta_{\,0\,0}(\lambda,\,\theta)\,$. This gives us an algebraic equation to determine the required profile $\eta_{\,0\,0}(\lambda,\,\theta)\,$:
\begin{equation*}
  -\rho\,g\,(R\ +\ \eta_{\,0\,0})\ +\ \frac{\rho\,(R\ +\ \eta_{\,0\,0})^2\,\Omega^{\,2}}{2}\;\sin^2\theta\ +\ C\ =\ 0\,.
\end{equation*}
The constant $C$ is determined from the condition that on the (North) pole $\theta\ =\ 0$ the free surface elevation is fixed. For simplicity we choose $\eta_{\,0\,0}(\lambda,\,0)\ =\ 0\,$. Then, the constant $C$ can be readily computed by evaluating the equation above at the North pole:
\begin{equation*}
  C\ =\ \rho\,g\,R\,.
\end{equation*}
With this value of $C$ in hands, the algebraic equation to determine the function $\eta_{\,0\,0}(\lambda,\,\theta)$ simply becomes:
\begin{equation*}
  -g\,\eta_{\,0\,0}\ +\ \frac{(R\ +\ \eta_{\,0\,0})^2\,\Omega^{\,2}}{2}\;\sin^2\theta\ =\ 0\,.
\end{equation*}
The physical sense has the following solution to the last equation:
\begin{equation}\label{eq:eta00}
  \eta_{\,0\,0}(\theta)\ =\ \frac{2}{g}\;\Omega^{\,2}\,R^2\,\sin^2\theta\cdot\Biggl[\,1\ +\ \sqrt{1\ -\ \frac{2}{g}\;\Omega^{\,2}\,R\,\sin^2\theta}\,\Biggr]^{\,-2}\,.
\end{equation}
This solution is represented in Figure~\ref{fig:sphere}(1). This solution will be used below as zero level $y\ =\ 0$ in free surface flows on globally flat geometries (see, for example, Part~I of the present series of papers \cite{Khakimzyanov2016c}). For example, the solid impermeable bottom of constant depth $h_0$ is given by the following equation
\begin{equation*}
  r\ =\ R\ +\ \eta_{\,0\,0}(\theta)\ -\ h_0\,.
\end{equation*}

\subsubsection{Boundary conditions}

When we model surface water waves, it is standard to use the full \textsc{Euler} equations as the governing equations. However, in the presence of impermeable bottom and free surface, the corresponding boundary conditions have to be specified \cite{Stoker1957}. From now on, we assume that the solid uneven moving bottom is given by the following equation:
\begin{equation*}
  r\ =\ R\ +\ \eta_{\,0\,0}(\theta)\ -\ h(t,\,\lambda,\,\theta)\ \defeq\ \rb(t,\,\lambda,\,\theta)\,,
\end{equation*}
and the free surface is given by
\begin{equation*}
  r\ =\ R\ +\ \eta_{\,0\,0}(\theta)\ +\ \eta(t,\,\lambda,\,\theta)\ \defeq\ \rs(t,\,\lambda,\,\theta)\,.
\end{equation*}
By taking the full (material) derivative of the last two equations with respect to time, we obtain two kinematic boundary conditions on the free surface and bottom respectively:
\begin{align}\label{eq:bc1}
  \eta_{\,t}\ +\ \V^{\,1}\,\eta_{\,\lambda}\ +\ \V^{\,2}\,(\eta\ +\ \eta_{\,0\,0})_{\,\theta}\ -\ \V^{\,3}\ &=\ 0\,, \qquad r\ =\ \rs\,, \\
  h_{\,t}\ +\ \V^{\,1}\,h_{\,\lambda}\ +\ \V^{\,2}\,(h\ -\ \eta_{\,0\,0})_{\,\theta}\ +\ \V^{\,3}\ &=\ 0\,, \qquad r\ =\ \rb\,. \label{eq:bc2}
\end{align}
Finally, on the free surface we also have the following dynamic condition:
\begin{equation*}
  \P\ =\ 0\,, \qquad r\ =\ \rs\,.
\end{equation*}
The last condition expresses the fact that the free surface is an isobar and the constant pressure is chosen to be zero without any loss of generality. Lateral boundary conditions are dependent on the application in hands and they have to be discussed separately.


\subsection{Modified Euler equations}
\label{sec:mEuler}

In order to derive shallow water equations in a moving curvilinear frame of reference, we have to estimate the relative importance of various terms already at the level of the full \textsc{Euler} equations \eqref{eq:cont}, \eqref{eq:moment}. Consequently, we have to pass to dimensionless variables. Let $\ell$ and $d$ be characteristic flow scales in horizontal and vertical directions correspondingly. Let $\alpha$ be the typical wave amplitude. Then, we can form three important dimensionless numbers:
\begin{description}
  \item[$\eps\ \eqdef\ \displaystyle{\frac{\alpha}{d}}\ $] Measure of the nonlinearity
  \item[$\mu\ \eqdef\ \displaystyle{\frac{d}{\ell}}\ $] Measure of the frequency dispersion
  \item[$\si\ \eqdef\ \displaystyle{\frac{d}{R}}\ $] Measure of flow thickness.
\end{description}
Parameters $\eps$ and $\mu$ are well known in long wave modelling (see, for example, \cite{Khakimzyanov2016c}), while parameter $\si$ is specific to globally spherical geometries. The values of all three parameters $(\eps,\,\mu,\,\si)$ characterize the aspect ratios of the flow. Various assumptions on the (relative) magnitude of these parameters allow to simplify more or less significantly the governing equations of Hydrodynamics.


\subsubsection{Dimensionless variables}

We shall use also some `derived' dimensionless quantities. Another dimensionless horizontal scale can be introduced on a sphere as
\begin{equation*}
  \ga\ \eqdef\ \frac{\si}{\mu}\ \equiv\ \frac{\ell}{R}\,.
\end{equation*}
The characteristic time scale $\tau$ is introduced as follows
\begin{equation*}
  \tau\ \eqdef\ \frac{\ell}{c}\,,
\end{equation*}
where $c\ \eqdef\ \sqrt{g\,d}$ is the usual (linear) gravity wave speed. Then, we can easily introduce the characteristic angular velocity of wave propagation:
\begin{equation*}
  \om\ \eqdef\ \frac{\ga}{\tau}\ \equiv\ \frac{c}{R}\,.
\end{equation*}
Finally, from now on we introduce also a new (independent) signed radial variable:
\begin{equation*}
  \ro\ \eqdef\ r\ -\ R\,.
\end{equation*}
Using the characteristic scales introduced above, we can scale all dependent and independent variables in our mathematical formulation:
\begin{equation*}
  \{\,\lambda^{\,\star},\,\theta^{\,\star}\,\}\ \ltimes\ \frac{\{\,\lambda,\,\theta\,\}}{\ga}\,, \quad
  \{\,h^{\,\star},\,\ro^{\,\star},\,\H^{\,\star}\,\}\ \ltimes\ \frac{\{\,h,\,\ro,\,\H\,\}}{d}\,, \quad \eta^{\,\star}\ \ltimes\ \frac{\eta}{\alpha}\,,
\end{equation*}
\begin{equation*}
  t^{\,\star}\ \ltimes\ \frac{t}{\tau}\,, \qquad
  \Omega^{\,\star}\ \ltimes\ \frac{\Omega}{\om}\,, \qquad
  \P^{\,\star}\ \ltimes\ \frac{\P}{\rho\,c^{\,2}}\,.
\end{equation*}
Contravariant components of the velocity vector are scaled as follows
\begin{equation*}
  \V^{\,\beta,\,\star}\ \ltimes\ \frac{\V^{\,\beta}}{\om}\,, \qquad
  \V^{\,3,\,\star}\ \ltimes\ \frac{\V^{\,3}}{\mu\,c}\,,
\end{equation*}
and covariant coordinates scale as
\begin{equation*}
  \V_{\,\beta}^{\,\star}\ \ltimes\ \frac{\V_{\,\beta}}{R\,c}\,, \qquad
  \V_{\,3}^{\,\star}\ \ltimes\ \frac{\V_{\,3}}{\mu\,c}\,,
\end{equation*}
where $\beta\ =\ 1,\,2\,$. Finally, we adimensionalize also the transformation \textsc{Jacobian} along with non-zero components of the covariant metric tensor:
\begin{equation*}
  \bigl\{\,\J^{\,\star},\,g_{\,1\,1}^{\,\star},\,g_{\,2\,2}^{\,\star}\,\bigr\}\ \ltimes\ \frac{\bigl\{\,\J,\,g_{\,1\,1},\,g_{\,2\,2}\,\bigr\}}{R^2}\,, \qquad
  g_{\,0\,1}^{\,\star}\ \ltimes\ \frac{g_{\,0\,1}}{R\,c}\,.
\end{equation*}


\subsubsection{Modification of Euler equations}

The parameter $\si$ represents the relative thickness of the liquid layer and usually in geophysical applications it rarely exceeds $\si\ \apprle\ 10^{-3}\,$. Consequently, the presence of this factor in front of $\O(1)$ terms shows their negligible importance. The modification of \textsc{Euler} equations consists in omitting such terms in equations \eqref{eq:seuler1}, \eqref{eq:seuler2}. Later, from these modified equations we shall derive the base wave model in Section~\ref{sec:nld}. The main difference between the original and modified \textsc{Euler} equations is that the \textsc{Jacobian} along with metric tensor components loose their dependence on the radial coordinate $r$ (or equivalently $\ro$) after the modification.

Consider, for example, the case of the \textsc{Jacobian} $\J$, which is computed using formula \eqref{eq:jacs} in the original \textsc{Euler} equations. In dimensionless variables the \textsc{Jacobian} becomes
\begin{multline*}
  \J^{\,\star}\ =\ \frac{\J}{R^2}\ =\ -\Bigl(\frac{r}{R}\Bigr)^2\,\sin\theta\ =\ -\Bigl(1\ +\ \frac{\ro}{R}\Bigr)^2\,\sin\theta\ =\\
  -\bigl(1\ +\ \si\,\ro^{\,\star}\bigr)^2\,\sin(\ga\,\theta^{\,\star})\ \simeq\ -\sin(\ga\,\theta^{\,\star})\,.
\end{multline*}
Above we neglected the term $\si\,\ro^{\,\star}\ =\ \O(\si)\,$. By making similar transformation with covariant metric tensor components, we obtain
\begin{equation*}
  g_{\,1\,1}^{\,\star}\ =\ \sin^2(\ga\,\theta^{\,\star})\,, \qquad
  g_{\,2\,2}^{\,\star}\ =\ 1\,, \qquad
  g_{\,0\,1}^{\,\star}\ =\ \Omega^{\,\star}\sin^2(\ga\,\theta^{\,\star})\,.
\end{equation*}
For the sake of completeness we provide also the modified quantities in dimensional variables as well:
\begin{equation}\label{eq:metric}
  \J\ =\ -R^2\,\sin\theta\,, \qquad
  g_{\,1\,1}\ =\ R^{\,2}\,\sin^2\theta\,, \qquad
  g_{\,2\,2}\ =\ R^{\,2}\,, \qquad
  g_{\,0\,1}\ =\ \Omega\,R^{\,2}\,\sin^2\theta\,.
\end{equation}
After these modification we obtain the modified \textsc{Euler} equations, which have the same form as \eqref{eq:seuler1}, \eqref{eq:seuler2}, but the \textsc{Jacobian} $\J$ and metric tensor $\{g_{\,i\,j}\}$ are modified as we explained hereinabove. The right-hand side in momentum equation \eqref{eq:seuler2} is modified as well $\S_{\,\beta}\ \rightsquigarrow\ \St_{\,\beta}$, since the quantities $g_{\,1\,1}$ and $g_{\,2\,2}$ do not depend on $r$ anymore:
\begin{equation*}
  \St_{\,\beta}\ =\ \begin{dcases}
    \ 0\,, & \beta\ =\ 1\,, \\
    \ (\Omega + \V^{\,1})^2\,R^{\,2}\,\sin\theta\,\cos\theta\,, & \beta\ =\ 2\,, \\
    \ -g\,, & \beta\ =\ 3\,.
  \end{dcases}
\end{equation*}


\subsubsection{Modified stationary free surface profile}

After the modifications we introduced into \textsc{Euler} equations, we have to reconsider accordingly the question of the stationary water profile, which will serve us as the unperturbed water level. The modified \textsc{Euler} equations for the quiescent fluid are
\begin{align*}
  \pd{\P}{\lambda}\ &=\ 0\,, \\
  \pd{\P}{\theta}\ &=\ \rho\,\Omega^{\,2}\,R^{\,2}\,\sin\theta\,\cos\theta\,, \\
  \pd{\P}{r}\ &=\ -\rho\,g\,.
\end{align*}
The last equations can be similarly integrated and the most general solution is
\begin{equation*}
  \P\ =\ -\rho\,g\,r\ +\ \frac{\rho\,\Omega^{\,2}\,R^{\,2}}{2}\;\sin^2\theta\ +\ C\,,
\end{equation*}
where $C\ =\ \const\ \in\ \R\,$. Imposing the dynamic boundary condition $\P(\lambda,\,\theta,\,r)\ =\ 0$ at the free surface $r\ =\ R\ +\ \eta_{\,0\,0}(\lambda,\,\theta)$ and the additional condition on the North pole $\eta_{\,0\,0}(\lambda,\,0)\ =\ 0\,$, we obtain the following expression for the required function $\eta_{\,0\,0}(\lambda,\,\theta)\,$:
\begin{equation}\label{eq:eta00m}
  \eta_{\,0\,0}(\lambda,\,\theta)\ \equiv\ \eta_{\,0\,0}(\theta)\ =\ \frac{\Omega^{\,2}\,R^{\,2}}{2\,g}\;\sin^2\theta\,, \qquad
  0\ \leq\ \theta\ \leq\ \pi\,.
\end{equation}
It can be readily seen that the equilibrium free surface profile $\eta_{\,0\,0}(\theta)$ predicted by modified \textsc{Euler} equations is different from the expression \eqref{eq:eta00} derived above. The modified expression \eqref{eq:eta00m} is depicted in Figure~\ref{fig:sphere}(2). However, it is not difficult to show that the modified expression \eqref{eq:eta00m} can be obtained by neglecting the terms of order $\O(\si)$ in the dimensionless counterpart of equation \eqref{eq:eta00}.

In case of modified \textsc{Euler} equations the bottom $r\ =\ \rb$ and free surface $r\ =\ \rs$ are described using corresponding deviations $-h(\lambda,\,\theta,\,t)$ and $\eta(\lambda,\,\theta,\,t)$ from the unperturbed water level $r\ =\ R\ +\ \eta_{\,0\,0}(\theta)\,$. The kinematic boundary conditions on the free surface \eqref{eq:bc1} and bottom \eqref{eq:bc2} for modified \textsc{Euler} equations remain unchanged.


\subsubsection{Modified Euler equations in dimensionless variables}

In this Section we summarize the developments made so far and, thus, we provide explicitly the modified \textsc{Euler} equations in scaled variables. For the sake of simplicity, we drop the superscript $\star$, which denotes dimensionless quantities. The continuity equation \eqref{eq:seuler1} becomes:
\begin{equation*}
  \bigl[\,\J\,\V^{\,1}\,\bigr]_{\,\lambda}\ +\ \bigl[\,\J\,\V^{\,2}\,\bigr]_{\,\theta}\ +\ \bigl[\,\J\,\V^{\,3}\,\bigr]_{\,\ro}\ =\ 0\,.
\end{equation*}
Non-conservative momentum equations \eqref{eq:seuler2} are
\begin{equation*}
  \bigl(1\ +\ (\mu^2 - 1)\,\delta_{\,\beta}^{\,3}\bigr)\cdot\bigl[\,\V_{\,\beta,\,t}\ +\ \V^{\,1}\cdot\V_{\,\beta,\,\lambda}\ +\ \V^{\,2}\cdot\V_{\,\beta,\,\theta}\ +\ \V^{\,3}\cdot\V_{\,\beta,\,\ro}\,\bigr]\ +\ \P_{\,q^{\,\beta}}\ =\ \S_{\,\beta}\,,
\end{equation*}
where $\beta\ =\ 1,\,2,\,3$ and as usually $q^{\,1}\ \equiv\ \lambda\,$, $q^{\,2}\ \equiv\ \theta\,$, $q^{\,3}\ \equiv\ \ro\,$. The second index denotes the partial derivative operation, \ie $\V_{\,\beta,\,q^{\,\alpha}}\ \eqdef\ \pd{\V_{\,\beta}}{q^{\,\alpha}}\,$. Covariant and contravariant components of the velocity are related in the following way:
\begin{equation*}
  \V_{\,1}\ =\ g_{\,0\,1}\ +\ g_{\,1\,1}\cdot\V^{\,1}\,, \qquad
  \V_{\,2}\ =\ \V^{\,2}\,, \qquad
  \V_{\,3}\ =\ \V^{\,3}\,.
\end{equation*}
Dimensionless right-hand side $\S_{\,\beta}$ has the following components:
\begin{equation}\label{eq:rhsS}
  \S_{\,1}\ =\ 0\,, \qquad
  \S_{\,2}\ =\ \frac{\si}{\mu}\;(\Omega\ +\ \V^{\,1})^2\,\sin(\ga\theta)\,\cos(\ga\theta)\,, \qquad
  \S_{\,3}\ =\ -1\,.
\end{equation}
Dimensionless kinematic and dynamic boundary conditions take the following form:
\begin{align*}
  \eps\,\eta_{\,t}\ +\ \eps\,\V^{\,1}\cdot\eta_{\,\lambda}\ +\ \V^{\,2}\cdot(\eps\,\eta\ +\ \eta_{\,0\,0})_{\,\theta}\ -\ \V^{\,3}\ &=\ 0\,, \qquad \ro\ =\ \ros\,, \\
  h_{\,t}\ +\ \V^{\,1}\cdot h_{\,\lambda}\ +\ \V^{\,2}\cdot(h\ -\ \eta_{\,0\,0})_{\,\theta}\ +\ \V^{\,3}\ &=\ 0\,, \qquad \ro\ =\ \rob\,, \\
  \P\ &=\ 0\,, \qquad \ro\ =\ \ros\,,
\end{align*}
where we introduced the traces of the shifted (dimensionless) radial variable $\ro$ at the bottom and free surface correspondingly:
\begin{equation*}
  \rob\ \eqdef\ \eta_{\,0\,0}\ -\ h\,, \qquad
  \ros\ \eqdef\ \eta_{\,0\,0}\ +\ \eps\,\eta\,.
\end{equation*}
Finally, the dimensionless `still' water level is give by the following formula:
\begin{equation*}
  \eta_{\,0\,0}(\theta)\ =\ \frac{\Omega^{\,2}}{2}\;\sin^2(\ga\theta)\,.
\end{equation*}
Below we shall need also the gradients of this profile:
\begin{equation}\label{eq:2.16}
  \partial_{\,t}\,\eta_{\,0\,0}\ =\ 0\,, \qquad
  \partial_{\,\lambda}\,\eta_{\,0\,0}\ =\ 0\,, \qquad
  \partial_{\,\theta}\,\eta_{\,0\,0}\ =\ \frac{\si}{\mu}\;\Omega^{\,2}\,\sin(\ga\theta)\,\cos(\ga\theta)\,.
\end{equation}


\section{Nonlinear dispersive shallow water wave model}
\label{sec:nld}

In order to derive an approximate long wave model, we shall work with dimensionless modified \textsc{Euler} equations summarized in the preceding Section. Moreover, by analogy with the globally flat case \cite{Khakimzyanov2016c}, we would like to separate momentum equations into `horizontal', \ie tangential and `vertical', \ie radial components. The complete set of equations is given here:
\begin{align}\label{eq:Bcont}
  \divf\bigl[\,\J\,\U\,\bigr]\ +\ \bigl[\,\J\,\W\,\bigr]_{\,\ro}\ &=\ 0\,, \\
  \Vv_{t}\ +\ \bigl(\U\scal\grad\bigr)\Vv\ +\ \W\,\Vv_{\,\ro}\ +\ \grad\P\ &=\ \Ss\,, \label{eq:Bhor} \\
  \mu^2\,\bigl(\W_{\,t}\ +\ \U\scal\grad\W\ +\ \W\,\W_{\,\ro}\bigr)\ +\ \P_{\,\ro}\ &=\ -1\,, \label{eq:Bvert}
\end{align}
where vectors $\U\ \eqdef\ \bigl(\V^{\,1},\,\V^{\,2}\bigr)$ and $\Vv\ \eqdef\ \bigl(\V_{\,1},\,\V_{\,2}\bigr)$ are contravariant and covariant components of the `horizontal' velocity. Moreover, we have the following relation among them
\begin{equation}\label{eq:vg}
  \Vv\ =\ \Gg\ +\ \Gm\scal\U\,, \qquad
  \Gg\ \eqdef\ \begin{pmatrix}
    g_{\,0\,1} \\
    0
  \end{pmatrix}\,, \qquad
  \Gm\ \eqdef\ \begin{pmatrix}
    g_{\,1\,1} & 0 \\
    0          & 1
  \end{pmatrix}\,.
\end{equation}
The `vertical' component of velocity was denoted by $\W\ \eqdef\ \V^{\,3}$. On the right hand side we have vector\footnote{In the sequel by ${}^{\top}$ we denote the transposition operator of linear objects such as vectors and matrices.} $\Ss\ \eqdef\ {}^{\top}(\S_{\,1},\,\S_{\,2})\,$. The `horizontal' gradient operator $\grad\ \eqdef\ \bigl(\partial_{\,\lambda},\,\partial_{\,\theta}\bigr)$ and the associated `flat' divergence operator is:
\begin{equation*}
  \divf\U\ \eqdef\ \pd{\V^{\,1}}{\lambda}\ +\ \pd{\V^{\,2}}{\theta}\,.
\end{equation*}
We have the following relations similar to the flat case:
\begin{equation*}
  \U\scal\grad\W\ =\ \V^{\,1}\cdot\W_{\,\lambda}\ +\ \V^{\,2}\cdot\W_{\,\theta}\,, \qquad
  \bigl(\U\scal\grad\bigr)\Vv\ =\ {}^{\top}\bigl(\U\scal\grad\V_{\,1},\, \U\scal\grad\V_{\,2}\bigr)\,.
\end{equation*}
In order to write boundary conditions in a compact form similar to the flat case, we introduce two new functions:
\begin{equation*}
  \hd\ \eqdef\ h\ -\ \eta_{\,0\,0}\,, \qquad
  \eps\,\etad\ \eqdef\ \eps\,\eta\ +\ \eta_{\,0\,0}\,.
\end{equation*}
Finally, the boundary conditions in new variables become
\begin{align}\label{eq:bc11}
  \eps\,\etad_{\,t}\ +\ \eps\,\U\scal\grad\etad\ -\ \W\ &=\ 0\,, \qquad \ro\ =\ \eps\,\etad\,, \\
  \hd_{\,t}\ +\ \U\scal\grad\hd\ +\ \W\ &=\ 0\,, \qquad \ro\ =\ -\hd\,, \label{eq:bc22} \\
  \P\ &=\ 0\,, \qquad \ro\ =\ \eps\,\etad\,. \label{eq:bc33}
\end{align}


\subsection{Horizontal velocity approximation}

In long wave models we describe traditionally the flow using the total water depth $\H$ and some velocity $\u(t,\,\lambda,\,\theta)\ =\ \bigl(u^{\,1}(t,\,\lambda,\,\theta),\,u^{\,2}(t,\,\lambda,\,\theta)\bigr)$ variable, which is supposed to approximate the `horizontal' velocity $\U(t,\,\lambda,\,\theta,\,\ro)\,$. In weakly dispersive models we can assume that $\u$ approximates $\U$ up to the order $\O(\mu^2)$ in the dispersion parameter. Mathematically it can be written as
\begin{equation}\label{eq:approxU}
  \U(t,\,\lambda,\,\theta,\,\ro)\ =\ \u(t,\,\lambda,\,\theta)\ +\ \mu^2\,\Ud(t,\,\lambda,\,\theta,\,\ro)\,.
\end{equation}
Here by $\Ud(t,\,\lambda,\,\theta,\,\ro)\ =\ \bigl(\Uc_d^{\,1}(t,\,\lambda,\,\theta,\,\ro),\,\Uc_d^{\,2}(t,\,\lambda,\,\theta,\,\ro)\bigr)$ we denote the dispersive\footnote{We call this component dispersive, since it disappears from the equations if we take the dispersionless limit $\mu\ \to\ 0\,$.} component of the velocity field. For potential flows one can compute explicitly an approximation to $\Ud(t,\,\lambda,\,\theta,\,\ro)\,$. However, in the present derivation we do not adopt this simplifying assumption.

For example, in \cite{Kirby2013} the authors choose $\u(t,\,\lambda,\,\theta)$ to be the `horizontal' flow velocity computed on a certain surface $\ro\ =\ \ro\,(t,\,\lambda,\,\theta)\,$ which lies between\footnote{The moving boundaries can be included.} the bottom and free surface so that the following expression makes sense:
\begin{equation*}
  \u(t,\,\lambda,\,\theta)\ \eqdef\ \U\bigl(t,\,\lambda,\,\theta,\,\ro\,(t,\,\lambda,\,\theta)\bigr)\,.
\end{equation*}
In other works (see \eg \cite{Wu1981, Ertekin1986}) $\u(t,\,\lambda,\,\theta)$ is taken to be the depth-averaged `horizontal' velocity $\U(t,\,\lambda,\,\theta,\,\ro)$ of (modified) \textsc{Euler} equations.

Using relation \eqref{eq:vg} we can similarly write the following decomposition for the covariant velocity vector $\Vv(t,\,\lambda,\,\theta,\,\ro)$ as a sum of a component $\v(t,\,\lambda,\,\theta)$ independent from the `vertical' coordinate and a dispersive addition $\Vd\ =\ \bigl(\V_{d,\,1}(t,\,\lambda,\,\theta,\,\ro),\,\V_{d,\,2}(t,\,\lambda,\,\theta,\,\ro)\bigr)\,$:
\begin{equation}\label{eq:defVv}
  \Vv(t,\,\lambda,\,\theta,\,\ro)\ =\ \v(t,\,\lambda,\,\theta)\ +\ \mu^2\,\Vd(t,\,\lambda,\,\theta,\,\ro)\,,
\end{equation}
where as in \eqref{eq:vg} we have the following relations:
\begin{equation}\label{eq:3.42}
  \v\ =\ \Gg\ +\ \Gm\scal\u\,, \qquad
  \Vd\ =\ \Gm\scal\Ud\,.
\end{equation}
By integrating the representation \eqref{eq:approxU} over the water depth, we trivially obtain:
\begin{equation}\label{eq:ave}
  \frac{1}{\H}\;\int_{-\hd}^{\,\eps\,\etad}\,\U(t,\,\lambda,\,\theta,\,\ro)\;\ud\ro\ =\ \u(t,\,\lambda,\,\theta)\ +\ \mu^2\,\Uu(t,\,\lambda,\,\theta)\,,
\end{equation}
where the total water depth $\H(t,\,\lambda,\,\theta)$ is defined as
\begin{equation*}
  \H\ \eqdef\ \eps\,\etad\ +\ \hd\ \equiv\ \eps\,\eta\ +\ h\,.
\end{equation*}
We introduced another depth-averaged contravariant velocity variable:
\begin{equation*}
  \Uu(t,\,\lambda,\,\theta)\ =\ {}^{\top}\bigl(\Uc^{\,1}(t,\,\lambda,\,\theta),\,\Uc^{\,2}(t,\,\lambda,\,\theta)\bigr)\ \eqdef\ \frac{1}{\H}\;\int_{-\hd}^{\,\eps\,\etad}\,\Ud(t,\,\lambda,\,\theta,\,\ro)\;\ud\ro\,.
\end{equation*}
Similarly, one can introduce the depth-averaged covariant velocity component:
\begin{equation*}
  \Vu(t,\,\lambda,\,\theta)\ =\ {}^{\top}\bigl(\V^{\,1}(t,\,\lambda,\,\theta),\,\V^{\,2}(t,\,\lambda,\,\theta)\bigr)\ \eqdef\ \frac{1}{\H}\;\int_{-\hd}^{\,\eps\,\etad}\,\Vd(t,\,\lambda,\,\theta,\,\ro)\;\ud\ro\,\ \equiv\ \Gm\scal\Uu\,.
\end{equation*}
The last identity comes from the independence of metric tensor components from the radial (`vertical') coordinate $\ro$ in modified \textsc{Euler} equations.

In general, the vector field $\Uu(t,\,\lambda,\,\theta)$ is not uniquely defined, unless some additional simplifying assumptions are adopted. For the moment we shall keep this function arbitrary (as we did in the globally plane case \cite{Khakimzyanov2016c}) to derive the most general base long wave model. However, before this model can be applied to any particular situation, one has to express $\Uu$ in terms of other variables $\H(t,\,\lambda,\,\theta)$ and $\u(t,\,\lambda,\,\theta)\,$. In Physics such relations are usually called the \emph{closures} (see \eg \cite{Bestion1990, Umlauf2005}).


\subsection{Continuity equation and the radial velocity component}

Let us integrate the continuity equation \eqref{eq:Bcont} in variable $\ro$ over the total water depth:
\begin{equation*}
  \divf\biggl[\,\J\,\int_{-\hd}^{\,\eps\,\etad}\U\;\ud\ro\,\biggr]\ -\ \left.\J\,\bigl(\eps\,\,\U\scal\grad\etad\ -\ \W\bigr)\right|^{\,\ro\,=\,\eps\,\etad}\ -\ \left.\J\,\bigl(\U\scal\grad\hd\ +\ \W\bigr)\right|_{\,\ro\,=\,-\hd}\ =\ 0\,.
\end{equation*}
From the last identity using boundary conditions \eqref{eq:bc11}, \eqref{eq:bc22} along with equation \eqref{eq:ave}, we obtain the continuity equation on a sphere:
\begin{equation*}
  (\J\,\H)_{\,t}\ +\ \divf\bigl[\,\J\,\H\,\u\,\bigr]\ =\ -\mu^2\;\divf\bigl[\,\J\,\H\,\Uu\,\bigr]\,.
\end{equation*}
Finally, by using the curvilinear divergence operator definition \eqref{eq:divdef} we can rewrite the continuity equation in a more familiar form (to be compared with the flat case \cite{Khakimzyanov2016c}):
\begin{equation}\label{eq:contH}
  \H_{\,t}\ +\ \div\bigl[\,\H\,\u\,\bigr]\ =\ -\mu^2\,\div\bigl[\,\H\,\Uu\,\bigr]\,,
\end{equation}
where we remind that
\begin{equation*}
  \div\bigl[\,\H\,\u\,\bigr]\ \equiv\ \frac{(\J\,\H\,\u^{\,1})_{\,\lambda}\ +\ (\J\,\H\,\u^{\,2})_{\,\theta}}{\J}\,, \qquad
  \div\bigl[\,\H\,\Uu\,\bigr]\ \equiv\ \frac{(\J\,\H\,\Uc^{\,1})_{\,\lambda}\ +\ (\J\,\H\,\Uc^{\,2})_{\,\theta}}{\J}\,.
\end{equation*}

By integrating the same continuity equation \eqref{eq:Bcont} in the radial coordinate from $-\hd$ to $\ro$ we obtain
\begin{equation}\label{eq:approxW}
  \W\ =\ \underbrace{-\Dd\,\hd\ -\ (\ro\ +\ \hd)\,\div\u}_{\displaystyle(\blacklozenge)}\ +\ \O(\mu^2)\,.
\end{equation}
It is natural to adopt the term $(\blacklozenge)$ as the `vertical' velocity in the nonlinear dispersive model:
\begin{equation*}
  \w\ \eqdef\ -\Dd\,\hd\ -\ (\ro\ +\ \hd)\,\div\u\,, \qquad \w\ \preccurlyeq\ \W\ +\ \O(\mu^2)\,.
\end{equation*}
The last relation $\w\ \preccurlyeq\ \W$ denotes the fact that a quantity $\w$ is obtained from $\W$ by an asymptotic truncation. Thus, we can say informally that $\w$ contains less information than $\W\,$. The operator $\Dd$ is the `horizontal' material derivative defined traditionally as
\begin{equation*}
  \Dd\ \eqdef\ \pd{}{t}\ +\ \u\scal\grad\,.
\end{equation*}


\subsection{Pressure representation}

In order to derive the pressure field approximation in terms of variables $\H$ and $\u$, we integrate the `vertical' momentum equation \eqref{eq:Bvert} over the radial coordinate from $\ro$ to $\eps\,\etad$. By taking into account the `horizontal' velocity ansatz \eqref{eq:approxU} we obtain 
\begin{equation}\label{eq:pressint}
  \P\ =\ \mu^2\,\int_{\,\ro}^{\,\eps\,\etad}\underbrace{\bigl(\Dd\,\W\ +\ \W\,\W_{\rho}\ +\ \O(\mu^2)\bigr)}_{\displaystyle(\lozenge)}\;\ud\rho\ -\ \ro\ +\ \eps\,\etad\,.
\end{equation}
We transform the expression $(\lozenge)$ under the integral by employing the approximation \eqref{eq:approxW} for the `vertical' velocity component $\W\ \succcurlyeq\ \w\,$:
\begin{multline*}
  \Dd\,\W\ +\ \W\,\W_{\,\ro}\ =\ \Dd\,\w\ +\ \w\;\pd{\w}{\ro}\ +\ \O(\mu^2)\ =\\ -\Dd^{\,2}\,\hd\ -\ (\ro\ +\ \hd)\cdot\Dd\,(\div\u)\ +\ (\ro\ +\ \hd)\scal(\div\u)^2\ +\ \O(\mu^2)\ \equiv\\ -(\ro\ +\ \hd)\,\Rr_1\ -\ \Rr_2\ +\ \O(\mu^2)\,,
\end{multline*}
where we introduced
\begin{equation*}
  \Rr_1\ \eqdef\ \Dd\,(\div\u)\ -\ (\div\u)^2\,, \qquad
  \Rr_2\ \eqdef\ \Dd^{\,2}\,\hd\,.
\end{equation*}

By substituting the last asymptotic approximation for $(\lozenge)$ into \eqref{eq:pressint}, we obtain the approximate pressure distribution over the fluid layer:
\begin{equation*}
  \P\ =\ \underbrace{\H\ -\ (\ro\ +\ \hd)\ -\ \mu^2\biggl[\,\Bigl(\H\ -\ (\ro + \hd)\Bigr)\;\Rr_2\ +\ \Bigl(\,\frac{\H^{\,2}}{2}\ -\ \frac{(\ro + \hd)^2}{2}\,\Bigr)\;\Rr_1\,\biggr]}_{\displaystyle{(\Box)}}\ +\ \O(\mu^4)\,.
\end{equation*}
The term $(\Box)$ will serve us as the pressure in the nonlinear long wave model:
\begin{equation}\label{eq:press}
  \Pi\ =\ \H\ -\ (\ro\ +\ \hd)\ -\ \mu^2\biggl[\,\Bigl(\H\ -\ (\ro + \hd)\Bigr)\;\Rr_2\ +\ \Bigl(\,\frac{\H^{\,2}}{2}\ -\ \frac{(\ro + \hd)^2}{2}\,\Bigr)\;\Rr_1\,\biggr]\,.
\end{equation}
We would like to insist on two important facts that have been just shown:
\begin{itemize}
  \item Pressure $\Pi$ approximates the three-dimensional pressure distribution $\P$ to the asymptotic order $\O(\mu^4)$, \ie $\P\ =\ \Pi\ +\ \O(\mu^4)\,$. This fact we will be also denoted by $\Pi\ \preccurlyeq\ \P\ +\ \O(\mu^4)\,$.
  \item The expression \eqref{eq:press} does not depend on the variable $\Uu\,$, which is to be specified later.
\end{itemize}

\begin{remark}
One can notice that
\begin{equation*}
  \Dd^{\,2}\,\hd\ =\ \Dd^{\,2}\,h\ -\ \Dd^{\,2}\,\eta_{\,0\,0}\ =\ \Dd^{\,2}\,h\ -\ \Dd\,\Bigl(\frac{\si}{\mu}\;u^{\,2}\,\Omega^{\,2}\,\sin(\ga\theta)\,\cos(\ga\theta)\Bigr)\,.
\end{equation*}
Thus, we can write:
\begin{equation*}
  \Dd^{\,2}\,\hd\ =\ \D^{\,2}\,h\ +\ \O\Bigl(\,\frac{\si}{\mu}\,\Bigr)\,.
\end{equation*}
By noticing that terms $\Rr_{1,\,2}$ appear with the coefficient $\mu^2\,$, one can equivalently define $\Rr_2$ as
\begin{equation*}
  \Rr_2^{\,\star}\ \eqdef\ \D^{\,2}\,h\,,
\end{equation*}
consistently with modified \textsc{Euler} equations. The difference between two quantities being asymptotically negligible, \ie
\begin{equation*}
  \mu^2\,\bigl(\Rr_2\ -\ \Rr_2^{\,\star}\bigr)\ =\ \O(\si\,\mu)\,.
\end{equation*}
\end{remark}


\subsection{Momentum equations}

In order to derive the `horizontal' momentum equations in the nonlinear dispersive wave model, we integrate the horizontal momentum equation \eqref{eq:Bhor} over the fluid layer depth. By taking into account the dynamic boundary condition \eqref{eq:bc33}, we obtain:
\begin{equation}\label{eq:horint}
  \int_{\,-\hd}^{\,\eps\,\etad}\underbrace{\bigl(\Vv_{t}\ +\ \bigl(\U\scal\grad\bigr)\Vv\ +\ \W\,\Vv_{\,\ro}\bigr)}_{\displaystyle(\Maltese)}\;\ud\ro\ +\ \grad\int_{-\hd}^{\,\eps\,\etad}\P\;\ud\ro\ -\ \left.\P\,\right|_{\,\ro\,=\,-\hd}\cdot\grad\hd\ =\ \int_{\,-\hd}^{\,\eps\,\etad}\Ss\;\ud\ro\,.
\end{equation}
We underline the fact that the last identity is exact. In order to simplify this equation, we exploit first the approximation for $\P$ worked out above:
\begin{equation*}
  \grad\int_{\,-\hd}^{\,\eps\,\etad}\,\P\;\ud\ro\ =\ \grad\Pp\ +\ \O(\mu^4)\,, \qquad
  \left.\P\right|_{\,\ro\,=\,-\hd}\ =\ \pc\ +\ \O(\mu^4)\,,
\end{equation*}
where we introduced the depth-integrated and bottom pressures defined respectively as
\begin{align*}
  \Pp\ &\eqdef\ \int_{\,-\hd}^{\,\eps\,\etad}\,\Pi\;\ud\ro\ \equiv\ \frac{\H^{\,2}}{2}\ -\ \mu^2\,\Pnh\,, \\
  \pc\ &\eqdef\ \left.\Pi\,\right|_{\,\ro\,=\,-\hd}\ \equiv\ \H\ -\ \mu^2\,\pb\,.
\end{align*}
Above we separated hydrostatic terms from non-hydrostatic ones summarized in functions $\Pnh$ and $\pb$ defined as
\begin{equation*}
  \Pnh\ \eqdef\ \frac{\H^{\,3}}{3}\;\Rr_1\ +\ \frac{\H^{\,2}}{2}\;\Rr_2^{\,\star}\,, \qquad
  \pb\ \eqdef\ \frac{\H^{\,2}}{2}\;\Rr_1\ +\ \H\,\Rr_2^{\,\star}\,.
\end{equation*}
Physically, in our dispersive wave model $\pb$ is the non-hydrostatic pressure trace at the bottom and $\Pnh$ is the depth-integrated non-hydrostatic pressure component. Then, we have
\begin{equation*}
  \pb\,\grad\hd\ =\ \bigl(\H\ -\ \mu^2\,\pb\bigr)\cdot\bigl(\grad h\ -\ \grad\eta_{\,0\,0}\bigr)\ =\ \bigl(\H\ -\ \mu^2\,\pb\bigr)\,\grad h\ -\ \H\,\grad\eta_{\,0\,0}\ +\ \mu^2\,\pb\,\grad\eta_{\,0\,0}\,.
\end{equation*}
From \eqref{eq:2.16} it follows that
\begin{equation*}
  \grad\eta_{\,0\,0}\ =\ \begin{pmatrix}
    0 \\
    \dfrac{\si}{\mu}\;\Omega^{\,2}\,\sin(\ga\theta)\,\cos(\ga\theta)
  \end{pmatrix}\,,
\end{equation*}
and consequently we have
\begin{equation*}
  \pc\,\grad\hd\ =\ \bigl(\H\ -\ \mu^2\,\pb)\,\grad h\ -\ \H\,\grad\eta_{\,0\,0}\ +\ \O(\mu^4\ +\ \si\,\mu\ +\ \si^{\,2})\,.
\end{equation*}

Now we proceed to the approximation of remaining terms in the depth-integrated `horizontal' momentum equation \eqref{eq:horint}. The term involving the `vertical' velocity component can be approximated as follows
\begin{multline*}
  \int_{\,-\hd}^{\,\eps\,\etad}\,\W\,\Vv_{\,\ro}\;\ud\ro\ =\ \mu^2\,\int_{\,-\hd}^{\,\eps\,\etad}\,\w\;\pd{\Vd}{\ro}\;\ud\ro\ +\ \O(\mu^4)\ =\\ \mu^2\Bigl\{\,\left.(\w\,\Vd)\,\right|_{\,\ro\,=\,-\hd}^{\,\ro\,=\,\eps\,\etad}\ -\ \int_{\,-\hd}^{\,\eps\,\etad}\,\Vd\,\pd{\w}{\ro}\;\ud\ro\,\Bigr\}\ +\ \O(\mu^4)\ =\\ \mu^2\,\Bigl\{-\left.\bigl(\Dd\,\hd\ +\ \H\,(\div\u)\bigr)\,\Vd\,\right|^{\,\ro\,=\,\eps\,\etad}\ +\ \left.\Dd\,\hd\,\Vd\,\right|_{\,\ro\,=\,-\hd}\ +\ \H\,\Vu\,(\div\u)\Bigr\}\ +\ \O(\mu^4)\,.
\end{multline*}
The group of terms in \eqref{eq:horint} involving the `horizontal' velocities are similarly transformed:
\begin{multline*}
  \int_{\,-\hd}^{\,\eps\,\etad}\,\bigl[\,\Vv_{t}\ +\ \bigl(\U\scal\grad\bigr)\Vv\,\bigr]\;\ud\ro\ =\ \int_{\,-\hd}^{\,\eps\,\etad}\,\Bigl[\,\bigl(\v\ +\ \mu^2\,\Vd\bigr)_{\,t}\ +\ \Bigl(\bigl(\u\ +\ \mu^2\,\Ud\bigr)\scal\grad\Bigr)\cdot\bigl(\v\ +\ \mu^2\,\Vd\bigr)\,\Bigr]\;\ud\ro\ =\\
  \int_{\,-\hd}^{\,\eps\,\etad}\,\Bigl[\,\v_{\,t}\ +\ (\u\scal\grad)\,\v\,\Bigr]\;\ud\ro\ +\ \mu^2\,\int_{\,-\hd}^{\,\eps\,\etad}\,\Bigl[\,\Vd_{t}\ +\ (\u\scal\grad)\,\Vd\ +\ (\Ud\scal\grad)\,\v\,\Bigr]\;\ud\ro\ +\ \O(\mu^4)\ =\\
  \H\,\bigl[\,\v_{\,t}\ +\ (\u\scal\grad)\,\v\,\bigr]\ +\ \mu^2\;\Bigl\{\,(\H\,\Vu)_{\,t}\ -\ \eps\,\etad_{\,t}\,\Vd\,\bigr\rvert^{\,\ro\,=\,\eps\,\etad}\ -\ \hd_{\,t}\,\Vd\,\bigr\rvert_{\,\ro\,=\,-\hd}\ +\ (\u\scal\grad)\cdot(\H\,\Vu)\\
  -\ \eps\,(\u\scal\grad\etad)\,\Vd\,\bigr\rvert^{\,\ro\,=\,\eps\,\etad}\ -\ (\u\scal\grad\hd)\,\Vd\,\bigr\rvert_{\,\ro\,=\,-\hd}\ +\ \H\,(\Uu\scal\grad)\,\v\,\Bigr\}\ +\ \O(\mu^4)\ = \\
  \H\,\bigl[\,\v_{\,t}\ +\ (\u\scal\grad)\,\v\,\bigr]\ +\ \mu^2\,\Bigl\{\,(\H\,\Vu)_{\,t}\ +\ (\u\scal\grad)\,(\H\,\Vu)\ +\\
  \H\,(\Uu\scal\grad)\,\v\ -\ \eps\,(\Dd\,\etad)\,\Vd\,\bigr\rvert^{\,\ro\,=\,\eps\,\etad}\ -\ (\Dd\,\hd)\,\Vd\,\bigr\rvert_{\,\ro\,=\,-\hd}\,\Bigr\}\ +\ \O(\mu^4)\,.
\end{multline*}
Combining the last two results we obtain the following expression for the term $(\Maltese)$:
\begin{multline*}
  \int_{\,-\hd}^{\,\eps\,\etad}\,\underbrace{\bigl(\Vv_{t}\ +\ \bigl(\U\scal\grad\bigr)\Vv\ +\ \W\,\Vv_{\,\ro}\bigr)}_{\displaystyle{(\Maltese)}}\;\ud\ro\ =\ \H\,\bigl[\,\v_{\,t}\ +\ (\u\scal\grad)\,\v\,\bigr]\\ -\ \mu^2\,\underbrace{\bigl[\,\Dd\,\H\ +\ \H\,(\div\u)\,\bigr]}_{\displaystyle(\maltese)}\,\Vd\bigr\rvert^{\,\ro\,=\,\eps\,\etad}\\ +\ \mu^2\,\Bigl\{(\H\,\Vu)_{\,t}\ +\ (\u\scal\grad)\cdot(\H\,\Vu)\ +\ \H\,(\Uu\scal\grad)\,\v\ +\ \H\,\Vu\,(\div\u)\Bigr\}\ +\ \O(\mu^4)\,.
\end{multline*}
The term $(\maltese)$ can be simplified by taking into account the mass conservation equation \eqref{eq:contH}:
\begin{equation*}
  (\maltese)\ \equiv\ \Dd\,\H\ +\ \H\,(\div\u)\ \overset{\mbox{\eqref{eq:contH}}}{=}\ -\mu^2\,\div(\H\,\Uu)\ =\ \O(\mu^2)\ \succcurlyeq\ 0\,.
\end{equation*}
The last step consists in averaging the right hand side of equation \eqref{eq:Bhor}:
\begin{equation*}
  \int_{\,-\hd}^{\,\eps\,\etad}\,\Ss\;\ud\ro\ =\ \H\,\Sr\ +\ \O(\si\,\mu)\,.
\end{equation*}
The term $\O(\si\,\mu)$ can be consistently neglected in accordance with modified \textsc{Euler} equations. The components of $\Ss\ =\ {}^{\top}(\S_{\,1},\,\S_{\,2})$ are defined in \eqref{eq:rhsS}. The first component $\S_{\,1}\ \equiv\ 0$, so its averaging is rather a trivial task. Let us focus on $\S_{\,2}$ component:
\begin{equation*}
  \S_2\ =\ \frac{\si}{\mu}\;\bigl(\Omega\ +\ u^{\,1}\ +\ \mu^2\,\boldsymbol{U}_{d}^{1}\bigr)^2\;\sin(\ga\theta)\,\cos(\ga\theta)\ =\ \underbrace{\frac{\si}{\mu}\;\bigl(\Omega\ +\ u^{\,1}\bigr)^2\;\sin(\ga\theta)\,\cos(\ga\theta)}_{\displaystyle{\Sr_2}}\ +\ \O(\si\,\mu)\,.
\end{equation*}
Thus, $\Sr\ =\ {}^{\top}(0,\,\Sr_{\,2})$ with $\Sr_{\,2}$ defined above.

After combining all the developments made above and neglecting the terms of order $\O(\mu^4)$, we obtain the required momentum balance equation in dimensionless variables:
\begin{multline}\label{eq:mom}
  \H\,\bigl(\v_{\,t}\ +\ (\u\scal\grad)\,\v\bigr)\ +\ \grad\Bigl(\frac{\H^{\,2}}{2}\Bigr)\ =\ \H\,\grad h\ +\ \H\,\bigl(\Sr\ -\ \grad\eta_{\,0\,0}\bigr)\ +\\
  \mu^2\,\bigl(\grad\Pnh\ -\ \pb\,\grad h\bigr)\ -\mu^2\,\Bigl[\,(\H\,\Vu)_{\,t}\ +\ (\u\scal\grad)\,(\H\,\Vu)\ +\ \H\,(\Uu\scal\grad)\,\v\ +\ \H\,\Vu\,(\div\u)\,\Bigr]\,.
\end{multline}


\subsection{Base model in conservative form}

For theoretical and numerical analysis of the governing equations, it can be beneficial to recast equations in the conservative form. In order to achieve this goal, we have to introduce the tensorial product operation $\otimes\,$. For our modest purposes it is sufficient to define this operation on vectors in $\R^2\,$. Let us take a covariant vector $\valpha\ =\ (\alpha_{\,1},\,\alpha_{\,2})$ and a contravariant vector $\vbeta\ =\ (\beta^{\,1},\,\beta^{\,2})\,$. Then, their tensorial product is defined as
\begin{equation*}
  \valpha\ \otimes\ \vbeta\ \eqdef\ \begin{pmatrix}
    \alpha_{\,1}\,\beta^{\,1} & \alpha_{\,1}\,\beta^{\,2} \\
    \alpha_{\,2}\,\beta^{\,1} & \alpha_{\,2}\,\beta^{\,2}
  \end{pmatrix}\,.
\end{equation*}
The divergence of the $2-$tensor $\valpha\ \otimes\ \vbeta$ is a vector defined as
\begin{equation*}
  \div(\valpha\ \otimes\ \vbeta)\ \eqdef\ \begin{pmatrix}
    \div(\alpha_{\,1}\,\vbeta) \\
    \div(\alpha_{\,2}\,\vbeta)
  \end{pmatrix}\,,
\end{equation*}
or in component-wise form as
\begin{equation}\label{eq:otimes}
  \div(\valpha\ \otimes\ \vbeta)\ \eqdef\ \frac{1}{\J}\;\begin{pmatrix}
    \bigl(\J\,\alpha_{\,1}\,\beta^{\,1}\bigr)_{\,\lambda}\ +\ \bigl(\J\,\alpha_{\,1}\,\beta^{\,2}\bigr)_{\,\theta} \\
    \bigl(\J\,\alpha_{\,2}\,\beta^{\,1}\bigr)_{\,\lambda}\ +\ \bigl(\J\,\alpha_{\,2}\,\beta^{\,2}\bigr)_{\,\theta}
  \end{pmatrix}\,.
\end{equation}
Then, by using this definition of the tensor product (in curvilinear coordinates), one can show the following formula remains true by analogy with the usual (\ie `flat') vector calculus:
\begin{equation*}
  \div(\valpha\ \otimes\ \vbeta)\ \equiv\ \valpha\,(\div\vbeta)\ +\ (\vbeta\scal\grad)\,\valpha\,.
\end{equation*}
We have set up all the tools to transform the momentum equation \eqref{eq:mom}. By multiplying the continuity equation \eqref{eq:contH} by $\v$ and adding it to \eqref{eq:mom} we obtain the balance equation for the total `horizontal' momentum:
\begin{multline*}
  (\H\,\v)_{\,t}\ +\ \div(\H\,\v\,\otimes\,\u) \ +\ \grad\Bigl(\frac{\H^{\,2}}{2}\Bigr)\ =\ \H\,\grad h\ +\ \H\,\bigl(\Sr\ -\ \grad\eta_{\,0\,0}\bigr)\ +\\
  \mu^2\,\bigl(\grad\Pnh\ -\ \pb\,\grad h\bigr)\ -\mu^2\,\Bigl[\,(\H\,\Vu)_{\,t}\ +\ \div(\H\,\v\,\otimes\,\Uu)\ +\ \div(\H\,\Vu\,\otimes\,\u)\,\Bigr]\,.
\end{multline*}
By using formula \eqref{eq:otimes}, we can rewrite the last equation in partial derivatives:
\begin{multline*}
  (\J\,\H\,\v)_{\,t}\ +\ \bigl[\,\J\,\H\,\v\,u^{\,1}\,\bigr]_{\,\lambda}\ +\ \bigl[\,\J\,\H\,\v\,u^{\,2}\,\bigr]_{\,\theta}\ +\ \grad\Bigl(\J\;\frac{\H^{\,2}}{2}\Bigr)\ =\\ 
  \J\,\H\,\grad h\ +\ \frac{\H^{\,2}}{2}\;\grad\J\ +\ \J\,\H\,\bigl(\Sr\ -\ \grad\eta_{\,0\,0}\bigr)\ +\ 
  \mu^2\,\J\,\bigl(\grad\Pnh\ -\ \pb\,\grad h\bigr)\\ -\ \mu^2\,\Bigl\{\,(\J\,\H\,\Vu)_{\,t}\ +\ \bigl(\J\,\H\,(\Uc^{\,1}\,\v\ +\ u^{\,1}\,\Vu)\bigr)_{\,\lambda}\ +\ \bigl(\J\,\H\,(\Uc^{\,2}\,\v\ +\ u^{\,2}\,\Vu)\bigr)_{\,\theta}\,\Bigr\}\,.
\end{multline*}
The conservative structure of the governing equations will be exploited in the following Part~IV \cite{Khakimzyanov2016b} in order to construct a modern finite volume TVD scheme for the numerical simulation of nonlinear dispersive waves on a sphere.


\subsubsection{Transformation of the source term}

Some additional simplification can be achieved in the last equation if we analyze the expression $\J\,\H\,\bigl(\Sr\ -\ \grad\eta_{\,0\,0}\bigr)\,$. Indeed, this expression contains centrifugal force terms proportional to $\propto\ \Omega^{\,2}\,$. It is not difficult to see that these terms cancel each other due to the judicious choice of the `still water' level $r\ =\ \eta_{\,0\,0}(\theta)\,$. As a result, we obtain
\begin{equation*}
  \Sr\ -\ \grad\eta_{\,0\,0}\ =\ \begin{pmatrix}
    0 \\
    \dfrac{\si}{\mu}\;\bigl(2\,\Omega\,u^{\,1}\ +\ (u^{\,1})^2\bigr)\;\sin(\ga\theta)\,\cos(\ga\theta)
  \end{pmatrix}\ \defeq\ \Srs\,.
\end{equation*}
The momentum balance equation reads now
\begin{multline*}
  (\J\,\H\,\v)_{\,t}\ +\ \bigl[\,\J\,\H\,\v\,u^{\,1}\,\bigr]_{\,\lambda}\ +\ \bigl[\,\J\,\H\,\v\,u^{\,2}\,\bigr]_{\,\theta}\ +\ \grad\Bigl(\J\;\frac{\H^{\,2}}{2}\Bigr)\ =\\ 
  \J\,\H\,\grad h\ +\ \frac{\H^{\,2}}{2}\;\grad\J\ +\ \J\,\H\,\Srs\ +\ 
  \mu^2\,\J\,\bigl(\grad\Pnh\ -\ \pb\,\grad h\bigr)\\ -\ \mu^2\,\Bigl\{\,(\J\,\H\,\Vu)_{\,t}\ +\ \bigl(\J\,\H\,(\Uc^{\,1}\,\v\ +\ u^{\,1}\,\Vu)\bigr)_{\,\lambda}\ +\ \bigl(\J\,\H\,(\Uc^{\,2}\,\v\ +\ u^{\,2}\,\Vu)\bigr)_{\,\theta}\,\Bigr\}\,.
\end{multline*}


\subsection{Base model in dimensional variables}

In applications it can be useful to have the governing equations in dimensional (unscaled) variables. In this way we obtain the following system of equations which constitute what we call the \emph{base model} in the present study:
\begin{equation}\label{eq:pbase1}
  (\J\,\H)_{\,t}\ +\ \divf\bigl[\,\J\,\H\,\u\,\bigr]\ =\ -\divf\bigl[\,\J\,\H\,\Uu\,\bigr]\,,
\end{equation}
\begin{multline}\label{eq:pbase2}
  (\J\,\H\,\v)_{\,t}\ +\ \divf\bigl[\,\J\,\H\,\v\,\otimesf\,\u\,\bigr]\ +\ g\,\grad\Bigl(\J\;\frac{\H^{\,2}}{2}\Bigr)\ =\\ 
  g\,\H\,\J\,\grad h\ +\ \J\,\H\,\Srs\ +\ g\;\frac{\H^{\,2}}{2}\;\grad\J\ +\ \J\;\bigl(\,\grad\Pnh\ -\ \pb\,\grad h\,\bigr)\\
  -\ \Bigl\{\,(\J\,\H\,\Vu)_{\,t}\ +\ \bigl(\J\,\H\,(\Uc^{\,1}\,\v\ +\ u^{\,1}\,\Vu)\bigr)_{\,\lambda}\ +\ \bigl(\J\,\H\,(\Uc^{\,2}\,\v\ +\ u^{\,2}\,\Vu)\bigr)_{\,\theta}\,\Bigr\}\,,
\end{multline}
where $\otimesf$ is the usual `flat' tensorial product of two vectors, $\H\ \eqdef\ \eta\ +\ h$ and covariant $\v\ =\ {}^{\top}(v_{\,1},\,v_{\,2})\ $/ contravariant $\u\ =\ {}^{\top}(u^{\,1},\,u^{\,2})$ components of the velocity vectors are related as
\begin{equation*}
  \v\ =\ \Gg\ +\ \Gm\scal\u\,, \qquad
  \Gg\ \eqdef\ \begin{pmatrix}
    g_{\,0\,1} \\
    0
  \end{pmatrix}\,, \qquad
  \Gm\ \eqdef\ \begin{pmatrix}
    g_{\,1\,1} & 0 \\
    0          & g_{\,2\,2}
  \end{pmatrix}\,.
\end{equation*}
The \textsc{Jacobian} $\J$ and the metric tensor components $\{g_{\,i\,j}\}_{\,0\,\leq\,i,\,j\,\leq\,2}$ are computed as specified in \eqref{eq:metric}. The velocity variable $\Uu\ =\ {}^{\top}\bigl(\Uc^{\,1},\,\Uc^{\,2}\bigr)$ has to be specified by a closure relation. Then, the vector function $\Vu\ =\ {}^{\top}\bigl(\V^{\,1},\,\V^{\,2}\bigr)$ is recomputed using relation $\Vu\ =\ \Gm\scal\Uu\,$.

Physically, the vector function $\Uu$ describes the deviation of the chosen `horizontal' velocity variable $\u$ from the depth-averaged profile. Below we shall consider two particular (and also popular) choices of $\Uu$ leading to different models which can be already used in practical applications.


\section{Depth-averaged velocity variable}
\label{sec:depth}

One natural way to choose the approximate velocity variable $\u$ in the dispersive wave model is to take the three-dimensional `horizontal' velocity field $\U(t,\,\lambda,\,\theta,\,\ro)$ and replace it by its average over the total fluid depth, \ie
\begin{equation*}
  \u\ =\ \frac{1}{\H}\;\int_{\,-\hd}^{\,\etad}\,\U(t,\,\lambda,\,\theta,\,\ro)\;\ud\ro\,.
\end{equation*}
Then, from formula \eqref{eq:ave} it follows that $\Uu\ \equiv\ \vO\,$. It is probably the simplest possible closure relation that one can find. By substituting $\Uu\ =\ \vO$ into governing equations of the base model we obtain:
\begin{equation}\label{eq:4.2}
  (\J\,\H)_{\,t}\ +\ \divf\bigl[\,\J\,\H\,\u\,\bigr]\ =\ 0\,,
\end{equation}
\begin{multline}\label{eq:4.3}
  (\J\,\H\,\v)_{\,t}\ +\ \divf\bigl[\,\J\,\H\,\v\,\otimesf\,\u\,\bigr]\ +\ g\,\grad\Bigl(\J\;\frac{\H^{\,2}}{2}\Bigr)\ =\\ 
  g\,\H\,\J\,\grad h\ +\ \J\,\H\,\Srs\ +\ g\;\frac{\H^{\,2}}{2}\;\grad\J\
  +\ \J\;\bigl(\,\grad\Pnh\ -\ \pb\,\grad h\,\bigr)\,.
\end{multline}


\subsection{Base model in terms of the linear velocity components}
\label{sec:lin1}

The last system of equations describes the evolution of the total water depth $\H$ and two contravariant components $u^{\,1}\,$, $u^{\,2}$ of the velocity vector $\u\,$. However, for the numerical modelling and the interpretation of obtained results it can be more convenient to use directly the linear components $u\,$, $v$ of this vector:
\begin{equation*}
  u\ \eqdef\ \sqrt{g_{\,1\,1}}\;u^{\,1}\ \equiv\ R\,u^{\,1}\,\sin\theta\,, \qquad
  v\ \eqdef\ \sqrt{g_{\,2\,2}}\;u^{\,2}\ \equiv\ R\,u^{\,2}\,.
\end{equation*}
By using formulas \eqref{eq:metric} and relation \eqref{eq:3.42} equations \eqref{eq:4.2}, \eqref{eq:4.3} become:
\begin{equation}\label{eq:base1}
  (\H\,R\,\sin\theta)_{\,t}\ +\ \bigl[\,\H\,u\,\bigr]_{\,\lambda}\ +\ \bigl[\,\H\,v\,\sin\theta\,\bigr]_{\,\theta}\ =\ 0\,,
\end{equation}
\begin{multline}\label{eq:base2}
  (\H\,u\,R\,\sin\theta)_{\,t}\ +\ \Bigl[\,\H\,u^2\ +\ g\;\frac{\H^{\,2}}{2}\,\Bigr]_{\,\lambda}\ +\ \bigl[\,\H\,u\,v\,\sin\theta\,\bigr]_{\,\theta}\ =\ g\,\H\,h_{\,\lambda}\\
  -\H\,u\,v\,\cos\theta\ -\ \digamma\,\H\,v\,R\,\sin\theta\ +\ \Pnh_{\,\lambda}\ -\ \pb\,h_{\,\lambda}\,,
\end{multline}
\begin{multline}\label{eq:base3}
  (\H\,v\,R\,\sin\theta)_{\,t}\ +\ \bigl[\,\H\,u\,v\,\bigr]_{\,\lambda}\ +\ \Bigl[\,\bigl(\H\,v^2\ +\ g\;\frac{\H^{\,2}}{2}\bigr)\,\sin\theta\,\Bigr]_{\,\theta}\ =\ g\,\H\,h_{\,\theta}\,\sin\theta\\
  g\;\frac{\H^{\,2}}{2}\;\cos\theta\ +\ \H\,u^2\,\cos\theta\ +\ \digamma\,\H\,u\,R\,\sin\theta\ +\ \bigl(\Pnh_{\,\theta}\ -\ \pb\,h_{\,\theta}\bigr)\,\sin\theta\,,
\end{multline}
where $\digamma\ \eqdef\ 2\,\Omega\,\cos\theta$ is the \textsc{Coriolis} parameter defined in terms of the colatitude $\theta\,$. In equations above $u^2$ and $v^2$ denote $u\cdot u$ and $v\cdot v$ respectively. The quantities $\Rr_{\alpha}\,$, $\alpha\ =\ 1,\,2$ arising in $\Pnh$ and $\pb$ can be computed by the following formulas:
\begin{align*}
  \Rr_1\ &\eqdef\ (\div\u)_{\,t}\ +\ \frac{1}{R\,\sin\theta}\;\Bigl\{\,u\,(\div\u)_{\,\lambda}\ +\ v\,(\div\u)_{\,\theta}\,\sin\theta\,\Bigr\}\ -\ (\div\u)^2\,, \\
  \Rr_2^{\,\star}\ &\eqdef\ (\Dd\,h)_{\,t}\ +\ \frac{1}{R\,\sin\theta}\;\Bigl\{\,u\,(\Dd\,h)_{\,\lambda}\ +\ v\,(\Dd\,h)_{\,\theta}\,\sin\theta\,\Bigr\}\,,
\end{align*}
where the divergence operator $\div\u$ and the material derivative $\Dd\,h$ can be computed in linear velocity components $u\,$, $v$ as
\begin{align*}
  \div\u\ &\eqdef\ \frac{1}{R\,\sin\theta}\;\Bigl\{\,u_{\,\lambda}\ +\ (v\,\sin\theta)_{\,\theta}\,\Bigr\}\,, \\
  \Dd\,h\ &\eqdef\ h_{\,t}\ +\ \frac{1}{R\,\sin\theta}\;\Bigl\{\,u\,h_{\,\lambda}\ +\ v\,h_{\,\theta}\,\sin\theta\,\Bigr\}\,.
\end{align*}

The hydrodynamic model \eqref{eq:base1}--\eqref{eq:base3} presented in this Section is the base fully nonlinear weakly dispersive model with depth-averaged velocity written in dimensional variables. We stress out that in the derivation above the flow irrotationality has never been assumed. It plays the same r\^ole in the spherical geometry as the so popular nowadays \textsc{Serre}--\textsc{Green}--\textsc{Naghdi} equations \cite{Serre1953, Su1969, Green1974, Green1976} in the flat case (see \cite{Khakimzyanov2016c} for the derivation of the base model on a globally flat space). By applying further simplifications to these equations we can obtain weakly nonlinear dispersive and fully nonlinear dispersionless equations.


\subsection{Weakly nonlinear model}
\label{sec:wnm}

Above we considered the fully nonlinear base model with depth-averaged velocity variable. During the derivation of this model we have not assumed that the wave scaled amplitude $\eps\ =\ \O(1)$ is a small parameter. In the present Section we propose a weakly nonlinear weakly dispersive model. Analogues of this model have been used for the numerical modelling of tsunami propagation in the ocean \cite{Tkalich2007, Lovholt2008, Lovholt2010}.

Weakly nonlinear models can be easily obtained from their fully nonlinear counterparts by adopting the simplifying assumption $\eps\ \ll\ 1\,$. It is quite common to work in the so-called \textsc{Boussinesq} regime \cite{Boussinesq1877, DMII, Brocchini2013} where we relate the nonlinear parameter to the magnitude of the dispersion:
\begin{equation*}
  \eps\ =\ \O(\mu^2)\,.
\end{equation*}
Thus, the terms of order $\O(\eps^2\ +\ \eps\,\mu^2)$ can be neglected in the governing equations. As a result we obtain the same governing equations \eqref{eq:base1}--\eqref{eq:base3} with one important modification --- in the computation of dispersive terms we use the following linearized formulas\footnote{In the superscripts we use the abbreviation `wnl' which stands for `weakly nonlinear'.}:
\begin{align*}
  \Pnh\ &\eqdef\ \frac{h^{\,3}}{3}\;\Rr_{1}^{\,\mathrm{wnl}}\ +\ \frac{h^{\,2}}{2}\;\Rr_{2}^{\,\mathrm{wnl}}\,, \\
  \pb\ &\eqdef\ \frac{h^{\,2}}{2}\;\Rr_{1}^{\,\mathrm{wnl}}\ +\ h\;\Rr_{2}^{\,\mathrm{wnl}}\,,
\end{align*}
and $\Rr_{\alpha}^{\,\mathrm{wnl}}\,$, $\alpha\ =\ 1,\,2$ are defined as
\begin{equation*}
  \Rr_{1}^{\,\mathrm{wnl}}\ \eqdef\ (\div\u)_{\,t}\,, \qquad
  \Rr_{2}^{\,\mathrm{wnl}}\ \eqdef\ (\Dd h)_{\,t}\ +\ \frac{1}{R\,\sin\theta}\;\Bigl\{\,u\,h_{\,t\,\lambda}\ +\ v\,h_{\,t\,\theta}\,\sin\theta\,\Bigr\}\,.
\end{equation*}


\subsection{Dispersionless shallow water equations}

Another important particular system can be trivially obtained from the base model \eqref{eq:base1}--\eqref{eq:base3} with the depth-averaged velocity by neglecting the non-hydrostatic terms $\Pnh\,$, $\pb\,$, which have the asymptotic order $\O(\mu^2)\,$. We reiterate on the fact that the three-dimensional flow irrotationality is not needed as a simplifying assumption. In this way we obtain a dispersionless model similar to nonlinear shallow water (or \textsc{Saint}-\textsc{Venant}) equations in the globally flat space \cite{SV1871}. This system of equations \eqref{eq:base1}--\eqref{eq:base3} (without non-hydrostatic terms) has the hyperbolic type. Consequently, it is natural to use finite volume methods for the numerical discretization of these equations \cite{LeVeque1992, Barth2004}. This method was proven to be very successful in solving hydrodynamic problems in coastal areas (see \eg \cite{Dutykh2009a, Khakimzyanov2016d}). The unique form of equations \eqref{eq:base1}--\eqref{eq:base3} for the entire hierarchy of asymptotic hydrodynamic models is very beneficial for the development of efficient numerical algorithms. Namely, we expect that neglecting non-hydrostatic terms in the discrete equations will result in a robust finite volume scheme for the remaining hyperbolic part of the equations. Numerical discretizations respecting the hierarchy of hydrodynamic models will be developed in the following Part~IV \cite{Khakimzyanov2016b} by analogy with the globally flat space \cite{Khakimzyanov2016}.


\subsection{State-of-the-art}

In this Section we make a review of published literature devoted to the derivation and/or application of nonlinear dispersive wave models with the depth-averaged velocity variable. The case of the velocity defined on an arbitrary surface in the fluid bulk will be discussed below in Section~\ref{sec:surf} (see Section~\ref{sec:start} for the corresponding literature review).

We can report a \textsc{Boussinesq}-type model in spherical coordinates which uses the depth-averaged velocity variable \cite{Tkalich2007}. The equations presented in that study have the advantage of being written in the conservative form. The bathymetry is assumed to be stationary. The \textsc{Coriolis} force is taken into account. However, some nonlinear terms in the right hand side (such as $\H\,u\,v\,\cos\theta$ and $\H\,u^{\,2}\,\cos\theta$) are omitted. Finally, the dispersive terms are written as if the bottom were flat. In the notation of our study, \textsc{Dao} \& \textsc{Tkalich} (2007) take the dispersive terms as
\begin{equation*}
  \Pnh\ =\ \frac{\H\,h^2}{3}\;\Rr_{1}^{\,\mathrm{wnl}}\,, \qquad
  \pb\ \equiv\ 0\,, \qquad
  \Rr_{2}^{\,\mathrm{wnl}}\ \equiv\ 0\,.
\end{equation*}
Moreover, $\Rr_{1}^{\,\mathrm{wnl}}$ is simplified by assuming that $\cos\theta\ \ll\ 1\,$. For the justification of this form of dispersive terms \textsc{Dao} \& \textsc{Tkalich} refer to \cite{Horrillo2006}. Then, this weakly nonlinear and weakly dispersive model was incorporated into \texttt{TUNAMI-N2} code, which was used to study Sumatra 2004 event. The authors came to the conclusion that the inclusion of dispersive effects and Earth's sphericity are needed to reproduce the observed data. In the aforementioned paper \cite{Horrillo2006} the authors used \textsc{Cartesian} coordinates only. Their dispersive terms were directly transformed into spherical coordinates by \textsc{Dao} \& \textsc{Tkalich} (2007) \cite{Tkalich2007}. In the following publication \textsc{Horillo} \etal (2012) \cite{Horrillo2012} used the spherical coordinates and depth-integrated equations with non-hydrostatic pressure (similar non-hydrostatic barotropic models are well-known for the flat space, see \eg \cite{Casulli1999, Dawson2005}). However, in contrast to \cite{Horrillo2006}, in \cite{Horrillo2012} \textsc{Horillo} \etal do not write explicitly non-hydrostatic terms.

Another dispersive model with depth-averaged velocity variable in spherical geometry was published in \cite{Lovholt2008, Lovholt2010}. This weakly nonlinear and weakly dispersive model is presented in a non-conservative form. Their model is the spherical counterpart of the classical \textsc{Peregrine} model well-known in flat space \cite{Peregrine1967}. However, some additional dispersive terms are added in order to improve the linear dispersion properties. The new terms come with a coefficient $\gamma\,$. If we set $\gamma\ =\ 0$ in their model, one obtains weakly dispersive model presented above with all nonlinear dispersive and \textsc{Coriolis} terms neglected. Strictly speaking, only with $\gamma\ =\ 0$ their velocity variable can be interpreted as the depth-integrated one. For $\gamma\ \neq\ 0$ we rather have a spherical analog of \textsc{Beji}--\textsc{Nadaoka} system \cite{Beji1996}. The derivation of this model can be found only in a technical report \cite{Pedersen2008}. Later this model was called \textsc{GloBouss}. This model was applied to model the propagation of a trans-Atlantic hypothetic tsunami resulting from an eventual landslide at the \textsc{La Palma} island.

A linear dispersive model in spherical coordinates was used in \cite{Miyoshi2015}. The linearization was justified by the need to produce a `fast' solution for the operational real time tsunami hazard forecast. The spherical \textsc{Boussinesq} system was borrowed from \cite{Tanioka1999}. Recently a parallel implementation of spherical weakly nonlinear \textsc{Boussinesq} equations was reported in \cite{Baba2015}. The authors used a conservative form of equations along with conservative variables. The authors came to the following conclusions:
\begin{quote}
  \it
  [\,\dots\,] A clear discrepancy was apparent from comparison of tsunami waveforms derived from dispersive and non-dispersive simulations at the DART21418 buoy located in the deep ocean. Tsunami soliton fission near the coast recorded by helicopter observations was accurately reproduced by the dispersive model with the high-resolution grids [\,\dots\,]
\end{quote}

We are not aware of any fully nonlinear models using the depth-averaged velocity variable. In this respect the present work fills in this gap. In the following Part~IV \cite{Khakimzyanov2016b} we shall describe an efficient splitting\footnote{The splitting is naturally performed in the hyperbolic and elliptic parts of the governing equations.}-type approach to solve an important representative of this class of dispersive wave models.


\section{Velocity variable defined on a given surface}
\label{sec:surf}

Another hierarchy of nonlinear dispersive wave models can be obtained by making a different choice of the velocity variable. A practically important choice consists in taking the trace of the three-dimensional `horizontal' velocity field at a given surface lying in the fluid bulk, \ie
\begin{equation}\label{eq:defu}
  \u\,(\lambda,\,\theta,\,t)\ \eqdef\ \U\,\bigl(\lambda,\,\theta,\,\ro_{\sigma}\,(\lambda,\,\theta,\,t),\,t\bigr)\,.
\end{equation}
This choice was hinted in a pioneering paper by \textsc{Bona} \& \textsc{Smith} (1976) \cite{BS} and developed later by \textsc{Nwogu} (1993) \cite{Nwogu1993} in the flat case. In order to close the system, one has to specify also the `dispersive' component of the velocity field $\Ud\ =\ \Ud(\H,\,\u)$ in terms of other dynamic variables $\H$ and $\u\,$. With the choice of the velocity $\u$ as specified above in \eqref{eq:defu}, $\Ud\ \neq\ \vO$ in general (similar to the case described in Section~\ref{sec:depth}, where only the integral of $\Ud$ over the water column height has to vanish due to the choice of $\u$) and we need an additional assumption to close completely the system. Consequently, in this case we proceed as follows: first, we construct $\Ud$ to the required accuracy and then, we apply the depth-averaging operator to determine $\U\,$. For example, in \cite{Kirby2013} the authors assumed the 3D flow to be irrotational. Here we shall give a derivation under weaker assumptions. Namely, we assume that only first two components $\omega^{\,1,\,2}$ of the vorticity field \eqref{eq:vort} vanish. The `vertical' vorticity component $\omega^{\,3}$ can take any values. In dimensionless variables this assumption can be expressed as
\begin{equation}\label{eq:assume}
  \Vv_{\,\ro}\ =\ \mu^2\,\grad\W\,.
\end{equation}
By differentiating \eqref{eq:defVv} with respect to $\ro$ and substituting the last identity into asymptotic expansion \eqref{eq:approxW} we obtain
\begin{equation*}
  \pd{\Vd}{\ro}\ =\ \grad\,w\ +\ \O(\mu^2)\ =\ -\grad(\Dd\,\hd)\ -\ (\div\u)\,\grad\hd\ -\ (\ro\ +\ \hd)\,\grad(\div\u)\ +\ \O(\mu^2)\,.
\end{equation*}
By integrating this identity over the vertical coordinate $\ro$ one obtains the following expression for $\Vd\,$:
\begin{equation*}
  \Vd\ =\ -(\ro + \hd)\,\Bigl[\,\grad(\Dd\hd)\ +\ (\div\u)\,\grad\hd\,\Bigr]\ -\ \frac{(\ro + \hd)^{2}}{2}\;\grad(\div\u)\ +\ \Vd\bigr\vert_{\,\ro\,=\,-\hd}\ +\ \O(\mu^2)\,.
\end{equation*}
By using the connection between covariant and contravariant components $\Ud\ =\ \Gm^{-1}\cdot\Vd\,$, which follows from \eqref{eq:3.42}, we obtain an asymptotic approximation to the 3D `horizontal' velocity field:
\begin{equation}\label{eq:reprU}
  \U(\lambda,\,\theta,\,\ro,\,t)\ =\ \u(\lambda,\,\theta,\,t)\ +\ \mu^2\,\Gm^{-1}\cdot\underbrace{\biggl\{\,(\ro + \hd)\,\A\ +\ \frac{(\ro + \hd)^2}{2}\;\B\ +\ \Cs\,\biggr\}}_{\displaystyle{(\checkmark)}}\ +\ \O(\mu^4)\,,
\end{equation}
where we introduced three vectors:
\begin{align*}
  \A\ &\eqdef\ -\grad(\Dd\hd)\ -\ (\div\u)\,\grad\hd\,, \\
  \B\ &\eqdef\ -\grad(\div\u)\,, \\
  \Cs\ &\eqdef\ \Vd\bigr\vert_{\,\ro\,=\,-\hd}\,.
\end{align*}
In order to compute the term $\Cs$ in terms of other dynamic variables, we consider the asymptotic representation for $\U(\lambda,\,\theta,\,\ro,\,t)$ and evaluate \eqref{eq:reprU} at $\ro\ =\ \ro_{\sigma}(\lambda,\,\theta,\,t)\,$. By definition \eqref{eq:defu} we must have $\U(\lambda,\,\theta,\,\ro_{\sigma},\,t)\ \equiv\ \u\,$. Consequently, at $\ro\ =\ \ro_{\sigma}(\lambda,\,\theta,\,t)$ the expression in braces $(\checkmark)\ \equiv\ \vO$ must vanish. Thus, we obtain
\begin{equation*}
  \Cs\ =\ -(\ro_{\sigma} + \hd)\,\A\ -\ \frac{(\ro_{\sigma} + \hd)^2}{2}\;\B\,.
\end{equation*}
Thus, the substitution of this expression for $\Cs$ into \eqref{eq:reprU} yields
\begin{equation*}
  \U(\lambda,\,\theta,\,\ro,\,t)\ =\ \u(\lambda,\,\theta,\,t)\ +\ \mu^2\,\Gm^{-1}\cdot\biggl\{\,(\ro - \ro_{\,\sigma})\,\A\ +\ \frac{(\ro + \hd)^2\ -\ (\ro_{\sigma} + \hd)^2}{2}\;\B\,\biggr\}\ +\ \O(\mu^4)\,.
\end{equation*}
In other words, the distribution of the `horizontal' velocity is approximatively quadratic to the asymptotic order $\O(\mu^4)\,$. The last formula can be used to reconstruct approximatively the 3D velocity field by having in hands the solution of the dispersive system only. As another side result of the formula above we obtain easily the required expression for the dispersive correction $\Ud\,$:
\begin{equation*}
  \Ud\ =\ \Gm^{-1}\cdot\biggl\{\,(\ro - \ro_{\,\sigma})\,\A\ +\ \frac{(\ro + \hd)^2\ -\ (\ro_{\sigma} + \hd)^2}{2}\;\B\,\biggr\}\ +\ \O(\mu^2)\,.
\end{equation*}
By applying the depth-averaging operator we obtain also the required closure relation to close the base model:
\begin{equation*}
  \Uu\ =\ \Gm^{-1}\cdot\biggl\{\,\Bigl[\,\frac{\H}{2}\ -\ (\ro_{\sigma} + \hd)\,\Bigr]\;\A\ +\ \Bigl[\,\frac{\H^{\,2}}{6}\ -\ \frac{(\ro_{\sigma} + \hd)^2}{2}\,\Bigr]\;\B\,\biggr\}\ +\ \O(\mu^2)\,.
\end{equation*}
The base model in physical variables has the same expressions \eqref{eq:pbase1}, \eqref{eq:pbase2}, but vector functions $\Uu$ and $\Vu$ have to be set accordingly to the closure presented in this Section.


\subsection{Further simplifications}

Some expressions and equations above can be further simplified by noticing that
\begin{align*}
  \grad(\Dd\hd)\ &=\ \grad(\Dd h)\ -\ \grad(\Dd\eta_{\,0\,0})\ =\ \grad(\Dd h)\ +\ \O\Bigl(\frac{\si}{\mu}\Bigr)\,, \\
  \grad\hd\ &=\ \grad h\ -\ \grad\eta_{\,0\,0}\ =\ \grad h\ +\ \O\Bigl(\frac{\si}{\mu}\Bigr)\,.
\end{align*}
Thus, $\grad(\Dd\hd)$ and $\grad\hd$ can be asymptotically interchanged with $\grad(\Dd h)$ and $\grad h$ correspondingly since $\Uu$ and $\Vu$ always appear in equations with coefficient $\mu^2\,$. Consequently, we have
\begin{equation*}
  \mu^2\,\Uu\ =\ \mu^2\,\Gm^{-1}\cdot\biggl\{\,\Bigl[\,\frac{\H}{2}\ -\ (r_{\sigma} + h)\,\Bigr]\;\A^{\,\star}\ +\ \Bigl[\,\frac{\H^{\,2}}{6}\ -\ \frac{(r_{\sigma} + h)^2}{2}\,\Bigr]\;\B\,\biggr\}\ +\ \O(\mu^4\ +\ \si\,\mu\ +\ \si^{\,2})\,,
\end{equation*}
where $r_{\sigma}\ \equiv\ \ro_{\sigma}\ -\ \eta_{\,0\,0}\,$, with $-\,h\ \leq\ r_{\sigma}\ \leq\ \eps\,\eta$ and
\begin{equation*}
  \A^{\,\star}\ \eqdef\ -\,\grad(\Dd\,h)\ -\ (\div\u)\,\grad h\,.
\end{equation*}

Let us summarize the developments made so far in this Section. First, the `horizontal' fluid velocity was defined in equation \eqref{eq:defu}. Then, we made a simplifying assumption \eqref{eq:assume}, which allowed us to derive the following closure relation:
\begin{equation}\label{eq:5.8}
  \Uu\ =\ \Gm^{-1}\cdot\biggl\{\,\Bigl[\,\frac{\H}{2}\ -\ (r_{\sigma} + h)\,\Bigr]\;\A^{\,\star}\ +\ \Bigl[\,\frac{\H^{\,2}}{6}\ -\ \frac{(r_{\sigma} + h)^2}{2}\,\Bigr]\;\B\,\biggr\}\,.
\end{equation}
The base model with velocity choice \eqref{eq:defu} has the same form \eqref{eq:pbase1}, \eqref{eq:pbase2} in dimensional variables. The invariance of equations with respect to the choice of `horizontal' velocity is among the main advantages of our modelling approach.


\subsection{Base model in terms of the linear velocity components}

Similarly as we did in Section~\ref{sec:lin1}, we can recast the base model \eqref{eq:pbase1}, \eqref{eq:pbase2} in terms of the components of linear velocity $u$ and $v$ (and we refer to Section~\ref{sec:lin1} for their definition):
\begin{equation*}
  \bigl(\H\,R\,\sin\theta\bigr)_{\,t}\ +\ \bigl[\,\H\,u\,\bigr]_{\,\lambda}\ +\ \bigl[\,\H\,v\,\sin\theta\,\bigr]_{\,\theta}\ =\ -\Bigl\{\,(\H\,U)_{\,\lambda}\ +\ (\H\,V\sin\theta)_{\,\theta}\,\Bigr\}\,,
\end{equation*}
\begin{multline*}
  \bigl(\H\,u\,R\,\sin\theta\bigr)_{\,t}\ +\ \Bigl[\,\H\,u^2\ +\ g\;\frac{\H^{\,2}}{2}\,\Bigr]_{\,\lambda}\ +\ \Bigl[\,\H\,u\,v\,\sin\theta\,\Bigr]_{\,\theta}\ =\\ g\,\H\,h_{\,\lambda}\ -\ \H\,u\,v\,\cos\theta\ -\ \digamma\,\H\,v\,R\,\sin\theta\ +\ \Pnh_{\,\lambda}\ -\ \pb\,h_{\,\lambda}\\
  -\Bigl\{\,\bigl(\H\,U\,R\,\sin\theta\bigr)_{\,t}\ +\ \bigl(2\,\H\,U\,u\bigr)_{\,\lambda}\ +\ \bigl(\H\,(\,U\,v\ +\ V\,u\,\bigr)\,\sin\theta\bigr)_{\,\theta}\,\Bigr\}\,,
\end{multline*}
\begin{multline*}
  \bigl(\H\,v\,R\,\sin\theta\bigr)_{\,t}\ +\ \Bigl[\,\H\,u\,v\,\Bigr]_{\,\lambda}\ +\ \Bigl[\,\Bigl(\H\,v^2\ +\ g\;\frac{\H^{\,2}}{2}\Bigr)\,\sin\theta\,\Bigr]_{\,\theta}\ =\\
  g\,\H\,h_{\,\theta}\,\sin\theta\ +\ g\;\frac{\H^{\,2}}{2}\;\cos\theta\ +\ \H\,u^2\,\cos\theta\ +\ \digamma\,\H\,u\,R\,\sin\theta\ +\ \bigl(\Pnh_{\,\theta}\ -\ \pb\,h_{\,\theta}\bigr)\,\sin\theta\\
  -\Bigl\{\,\bigl(\H\,V\,R\,\sin\theta\bigr)_{\,t}\ +\ \bigl(\H\,U\,v\ +\ \H\,V\,u\bigr)_{\,\lambda}\ +\ \bigl(2\,\H\,V\,v\,\sin\theta\bigr)_{\,\theta}\,\Bigr\}
\end{multline*}
where $U\ \eqdef\ R\,\Uc^{\,1}\sin\theta\,$, $V\ \eqdef\ R\,\Uc^2\,$ and $\Uc^{\,1,\,2}$ were defined in \eqref{eq:5.8}. The computation of non-hydrostatic pressure contributions are computed precisely as it is explained in Section~\ref{sec:lin1}. In the second equation above we omitted intentionally in the right hand side three terms $\H\,(U\,v\ +\ V\,u)\,\cos\theta$ and $\digamma\,\H\,V\,R\,\sin\theta$ since in dimensionless form they have the asymptotic order $\O(\si\,\mu)\,$.

One can notice that equations \eqref{eq:base1}--\eqref{eq:base3} can be obtained from the last system by setting $\Uc^{\,1,\,2}\ \equiv\ 0$ or equivalently $U\ =\ V\ \equiv\ 0\,$. The analogue of these equations was obtained in \cite{Kirby2013}. However, the system presented above admits a more elegant form since it enjoys the quasi-conservative mathematical structure.


\subsection{Weakly nonlinear model}

In order to obtain a weakly nonlinear model with a velocity variable defined on an arbitrary surface, it is sufficient to simplify accordingly the system of equations given in the previous Section.  Namely, all nonlinearities in the dispersive terms are to be neglected. The first group of terms to be simplified consists in non-hydrostatic pressure corrections $\grad\Pnh\ -\ \pb\,\grad h\,$, which is present for any choice of the velocity variable $\u$ (or equivalently for any closure relation $\Uu\ =\ \Uu\,(\H,\,\u)\,$). This simplification was explained above in Section~\ref{sec:wnm}. 

The second group of terms contains the vector $\Uu\,$. They are present in both continuity and momentum conservation equations. Vector $\Uu$ appears always with dimensionless coefficient $\mu^2\,$. Taking into account the \textsc{Boussinesq} (\ie weakly nonlinear) regime and definition $\H\ =\ h\ +\ \eps\,\eta\,$, we obtain that instead of closure relation \eqref{eq:5.8}, we have to use consistently
\begin{equation}\label{eq:5.13}
  \Uu\ =\ \Uu_{\,0}\ +\ \Gm^{-1}\cdot\biggl[\,\Bigl(\frac{h}{2}\ +\ r_{\sigma}\Bigr)\Bigl(\grad(\u\scal\grad h)\ +\ (\div\u)\grad h\Bigr)\ -\ \Bigl(\frac{h^{\,2}}{6}\ -\ \frac{(r_{\sigma}\ +\ h)^{\,2}}{2}\Bigr)\,\grad(\div\u)\,\biggr]\,,
\end{equation}
where
\begin{equation*}
  \Uu_{\,0}\ \eqdef\ \Bigl(\frac{h}{2}\ +\ r_{\sigma}\Bigr)\,\Gm^{-1}\cdot\grad h_{\,t}\,.
\end{equation*}
In other words, since coefficients $\A^{\,\star}$ and $\B$ are linear in velocities, then it was sufficient to replace $\H$ by $h\,$. The same operation has to be performed consistently in the right hand sides as well:
\begin{equation*}
  \bigl(\H\,U\bigr)_{\,\lambda}\ +\ \bigl[\,\H\,V\,\sin\theta\,\bigr]_{\,\theta}\ \rightsquigarrow\ \bigl(h\,U\bigr)_{\,\lambda}\ +\ \bigl[\,h\,V\,\sin\theta\,\bigr]_{\,\theta}\,,
\end{equation*}
\begin{equation*}
  \bigl(\H\,U\,R\,\sin\theta\bigr)_{\,t}\ \rightsquigarrow\ \bigl(h\,U\,R\,\sin\theta\bigr)_{\,t}\,, \qquad
  \bigl(\H\,V\,R\,\sin\theta\bigr)_{\,t}\ \rightsquigarrow\ \bigl(h\,V\,R\,\sin\theta\bigr)_{\,t}\,.
\end{equation*}
The components $U$ and $V$ are computed as above
\begin{equation*}
  U\ =\ R\,\Uc^{\,1}\sin\theta\,, \qquad V\ =\ R\,\Uc^{\,2}\,,
\end{equation*}
with the only difference is that $\Uc^{\,1,\,2}$ are given by equation \eqref{eq:5.13}.

Similar transformations (in this case linearizations) $\H\ \rightsquigarrow\ h\,$, $U\ \rightsquigarrow\ U_{\,0}\,$, $V\ \rightsquigarrow\ V_{\,0}\,$ have to be done in the remaining terms as well:
\begin{align*}
  2\,\bigl[\,\H\,u\,U\,\bigr]_{\,\lambda}\ +\ \bigl[\,\H\,(U\,v\ +\ V\,u)\,\sin\theta\,\bigr]_{\,\theta}\ &\rightsquigarrow\ 2\,\bigl[\,h\,u\,U_{\,0}\,\bigr]_{\,\lambda}\ +\ \bigl[\,h\,(U_{\,0}\,v\ +\ V_{\,0}\,u)\,\sin\theta\,\bigr]_{\,\theta}\,, \\
  \bigl[\,\H\,(U\,v\ +\ V\,u)\,\bigr]_{\,\lambda}\ +\ 2\,\bigl[\,\H\,v\,V\,\sin\theta\,\bigr]_{\,\theta}\ &\rightsquigarrow\ \bigl[\,h\,(U_{\,0}\,v\ +\ V_{\,0}\,u)\,\bigr]_{\,\lambda}\ +\ 2\,\bigl[\,h\,v\,V_{\,0}\,\sin\theta\,\bigr]_{\,\theta}\,,
\end{align*}
where $U_{\,0}$ and $V_{\,0}$ are defined through components of the vector $\Uu_{\,0}$ as
\begin{equation*}
  U_{\,0}\ \eqdef\ R\,\Uc_{\,0}^{\,1}\sin\theta\,, \qquad V_{\,0}\ \eqdef\ R\,\Uc_{\,0}^{\,2}\,.
\end{equation*}

Obtained in this way weakly nonlinear model is a spherical analogue of well-known \textsc{Nwogu} system on the plane \cite{Nwogu1993}. If we vanish the dispersive velocity correction $\Uu\ \equiv\ \vO$ we shall obtain the spherical counterpart of the classical \textsc{Peregrine} system \cite{Peregrine1967}.


\subsection{State-of-the-art}
\label{sec:start}

Let us mention a few publications which report the derivation or use of dispersive wave models on a sphere with the velocity defined on an arbitrary surface lying in the fluid bulk. A detailed derivation of the fully nonlinear weakly dispersive wave model with this choice of the velocity is given in \cite{Shi2012a, Kirby2013}. However, the resulting equations turn out to be cumbersome and in numerical simulations the Authors employ only the weakly nonlinear spherical \textsc{Boussinesq}-type equations. For example, in \cite{Harris2012} a numerical coupling between 3D \textsc{Navier}--\textsc{Stokes} (for the landslide area) and 2D spherical \textsc{Boussinesq} (for the far field propagation) is reported.

To our knowledge, the fully nonlinear model derived in the present study and earlier in \cite{Shi2012a} has never been used for large scale numerical simulations. It can be partially explained by the complexity of equations and by the lack of robust and efficient numerical discretizations for dispersive PDEs on a sphere.


\section{Discussion}
\label{sec:disc}

After the developments presented hereinabove in a globally spherical geometry, we finish the present manuscript by outlining the main conclusions and perspectives of our study.


\subsection{Conclusions}

In this work we derived a generic weakly dispersive but fully nonlinear model on a rotating, possibly deformed, sphere. This model contains a free contravariant $1-$tensor variable $\Uu$ (or its covariant equivalent $\Vu$). In order to close the system, one has to specify $\Uu$ as a function of two other model variables, \ie $\Uu\ =\ \Uu\,(\H,\,\u)\,$. This functional dependence is called the \emph{closure} relation. By choosing various closures, we show how one can obtain from the base model by simple substitutions some well-known models (or, at least, their fully nonlinear counterparts). Other choices of closure $\Uu\ =\ \Uu\,(\H,\,\u)$ lead to completely new models, whose properties are to be studied separately. Moreover, for any choice of the closure relation the base model has a nice conservative structure. Thus, this work can be considered as an effort towards further systematization of dispersive wave models on a spherical geometry. Moreover, the governing equations of the base model are given for the convenience in terms of the covariant/contravariant and linear velocity variables. For every model we give also its weakly nonlinear counterparts in case simpler models are needed. Of course, the classical nonlinear shallow water or \textsc{Saint}-\textsc{Venant} equations on a rotating sphere can be simply obtained from the base model by neglecting all non-hydrostatic terms.

Two popular closures were proposed in our study. In our study we always tried to use only the minimal assumptions about the three-dimensional flow. For instance, in contrast to \cite{Shi2012a, Kirby2013} we do not assume the flow to be irrotational. We note also the fact that the bottom was assumed to be unstationary. It allows to model tsunami generation by seismic \cite{Dutykh2007a, Dutykh2006, Dutykh2007b} and landslide \cite{Beisel2012, Dutykh2011d} mechanisms.


\subsection{Perspectives}

In the present manuscript the base model derivation was presented in a spherical geometry. This choice was made by the Authors due to the importance of applications in Meteorology, Climatology and Oceanography on global planetary scales. However, in this study we prepare a setting which could be fruitfully used in more general geometries. For instance, we believe that the techniques presented in this manuscript could be used to derive long wave models for shallow flows on compact manifolds. Curvilinear coordinates are routinely used in Fluid Mechanics. However, we believe that the right setting to work with Fluid Mechanics equations in complex geometries is the \textsc{Riemannian} geometry. In future studies we plan to show successful applications of this technology on more general geometries.

In the following (and the last) Part~IV \cite{Khakimzyanov2016b} of this series of papers we shall discuss the numerical discretization of nonlinear long wave models on globally spherical geometries. Namely, for the sake of simplicity, we shall take a particular \emph{avatar} of the base model and we shall show how to discretize it using modern finite volume schemes. After a direct generalization it can be easily extended to the base model as well, if it is needed, of course. The numerical solution of fully nonlinear dispersive wave equations with the velocity defined on an arbitrary level in the fluid bulk still constitutes a challenging problem which will be addressed in our future studies.


\subsection*{Acknowledgments}
\addcontentsline{toc}{subsection}{Acknowledgments}

This research was supported by RSCF project No 14--17--00219.


\appendix
\section{Reminder of basic tensor analysis}
\label{app:tensor}

In this Appendix (as well as in our study above) we adopt \textsc{Einstein}'s summation convention, \ie the summation is performed over repeating lower and upper indices. Moreover, indices denoted with Latin letters $i$, $j$, $k$, \etc~ vary from $0$ to $3$, while indices written with Greek letters $\alpha$, $\beta$, $\gamma$, \etc~ vary from $1$ to $3\,$. Some indices will be supplied with a prime, \eg $\alpha$, $\alpha^{\,\prime}$. In this case $\alpha$ and $\alpha^{\,\prime}$ should be considered as independent indices. Along with the diffeomorphism
\begin{equation}\label{eq:direct}
  x^0\ =\ q^0\ =\ t\,, \qquad
  x^\alpha\ =\ x^\alpha\;\bigl(q^0,\,q^1,\,q^2,\,q^3\bigr)\,, \qquad
  \alpha\ =\ 1,\,2,\,3\,,
\end{equation}
we shall consider also the inverse transformation of coordinates:
\begin{equation}\label{eq:inverse}
  q^0\ =\ x^0\ =\ t\,, \qquad
  q^\alpha\ =\ q^\alpha\;\bigl(x^0,\,x^1,\,x^2,\,x^3\bigr)\,, \qquad
  \alpha\ =\ 1,\,2,\,3\,.
\end{equation}
For the sake of convenience we introduce also short-hand notations for partial derivatives of the direct \eqref{eq:direct} and inverse transformations \eqref{eq:inverse}:
\begin{equation*}
  \D_{\,i}^{\ip}\ \eqdef\ \pd{x^{\ip}}{q^{\,i}}\,, \qquad
  \D_{\ip}^{\,i}\ \eqdef\ \pd{q^{\,i}}{x^{\ip}}\,,
\end{equation*}
which are equivalent to the series of definitions
\begin{align}\label{eq:1.6}
  \D_{\,0}^{\Op}\ =\ 1\,, \quad \D_{\alpha}^{\Op}\ =\ 0\,, \quad \D_{\,0}^{\ap}\ =\ \pd{x^{\ap}}{t}\, \quad \D_{\,\alpha}^{\ap}\ =\ \pd{x^\ap}{q^{\,\alpha}}\,,\\
  \D_{\Op}^{\,0}\ =\ 1\,, \quad \D_{\ap}^{\,0}\ =\ 0\,, \quad \D_{\Op}^{\,\alpha}\ =\ \pd{q^{\,\alpha}}{t}\, \quad \D_{\ap}^{\,\alpha}\ =\ \pd{q^{\,\alpha}}{x^{\ap}}\,.\label{eq:1.7}
\end{align}
We also have the following obvious identities:
\begin{equation}\label{eq:inv}
  \D_{\,\ip}^{\,i}\cdot\D_{\,j}^{\,\ip}\ =\ \delta_{\,j}^{\,i}\,, \qquad
  \D_{\,i}^{\,\ip}\cdot\D_{\,\jp}^{\,i}\ =\ \delta_{\,\jp}^{\,\ip}\,,
\end{equation}
where $\delta_{\,i}^{\,j}$ is the \textsc{Kronecker} symbol. We remind again that in the first formula there is an implicit summation over index $\ip$ and over $i$ in the second one.

In the sequel we complete the set of \textsc{Cartesian} basis vectors $\{\i_{\,\alpha}\}_{\alpha\,=\,1}^{3}$ with an additional vector $\i_{\,0}\ =\ (1,\,0,\,0,\,0)\,$. In other words, we use the standard orthonormal basis $\{\i_{\,i}\}_{i\,=\,0}^{3}$ in the \textsc{Euclidean} space $\R^4\ =\ \bigl\{(t,\,x^1,\,x^2,\,x^3)\bigr\}\,$. Sometimes, in order to introduce the summation, we shall equivalently employ basis vectors $\{\i^{\,i}\}_{i\,=\,0}^{3}$ with the upper index notation. 

The transformation of coordinates \eqref{eq:direct} along with the inverse transformation \eqref{eq:inverse} induces two new bases in $\R^4$:
\begin{description}
  \item[$\{\e_{\,i}\}_{i\,=\,0}^{3}\ $] moving \emph{covariant} basis
  \item[$\{\e^{i}\}_{i\,=\,0}^{3}\ $] moving \emph{contravariant} basis
\end{description}
By definition, the components of covariant and contravariant basis vectors can be expressed as follows:
\begin{equation*}
  \e_{\,i}\ \equiv\ \D_{\,i}^{\ip}\cdot\i_{\ip}\,, \qquad
  \e^{i}\ \equiv\ \D_{\ip}^{\,i}\cdot\i^{\ip}\,, \qquad
  i\ =\ 0,\,\ldots,\,3\,.
\end{equation*}
Right from this definition and relations \eqref{eq:inv} we can write down the connection among all the bases introduced so far:
\begin{equation*}
  \i_{\ip}\ =\ \D_{\ip}^{\,i}\,\e_i\ =\ \i^{\ip}\ =\ \D_{\,i}^{\ip}\,\e^{\,i}\,,
\end{equation*}
\begin{equation*}
  \e_{i}\ =\ \delta_{\ip\jp}\cdot\D_{\,i}^{\ip}\D_{\,j}^{\jp}\,\e^{\,j}\,, \qquad
  \e^{i}\ =\ \delta^{\ip\jp}\cdot\D^{\,i}_{\ip}\D^{\,j}_{\jp}\,\e_{\,j}\,, \qquad
  \e_{\,i}\scal\e^{\,j}\ =\ \delta_{\,i}^{\,j}\,.
\end{equation*}
Using vectors of these new bases we can define covariant $\{g_{\,ij}\}_{i,\,j\,=\,0}^{3}$ and contravariant $\{g^{\,ij}\}_{i,\,j\,=\,0}^{3}$ components of the metric tensor as scalar products of corresponding basis vectors:
\begin{align*}
  g_{\,ij}\ &\eqdef\ \e_i\scal\e_j\ \equiv\ \D_{\,i}^{\ip}\D_{\,j}^{\jp}\cdot\delta_{\ip\jp}\ \equiv\ g_{\,ji}\,, \\
  g^{\,ij}\ &\eqdef\ \e^i\scal\e^{\,j}\ \equiv\ \D_{\ip}^{\,i}\D_{\jp}^{\,j}\cdot\delta^{\ip\jp}\ \equiv\ g^{\,ji}\,.
\end{align*}
Moreover, thanks to \eqref{eq:inv}, one can easily check that
\begin{equation}\label{eq:inverse}
  g_{\,ij}\cdot g^{\,jk}\ =\ \delta_{\,i}^{\,k}\,.
\end{equation}
Using formulas \eqref{eq:1.6}, \eqref{eq:1.7} we obtain that
\begin{align}\label{eq:1.14}
  g_{\,00}\ =\ 1\ +\ \D_{\,0}^{\ap}\,\D_{\,0}^{\ap}\,\delta_{\ap\ap}\,, \quad
  g_{\,0\alpha}\ =\ \D_{\,0}^{\ap}\,\D_{\,\alpha}^{\ap}\,\delta_{\ap\ap}\,, \quad
  g_{\,\alpha\beta}\ =\ \D_{\,\alpha}^{\ap}\,\D_{\,\beta}^{\ap}\,\delta_{\ap\ap}\,, \\
  g^{\,00}\ =\ 1\,, \quad
  g^{\,0\alpha}\ =\ \D_{\Op}^{\,\alpha}\,, \quad
  g^{\,\alpha\beta}\ =\ \D_{\Op}^{\,\alpha}\,\D_{\Op}^{\,\beta}\ +\ \D_{\ap}^{\,\alpha}\,\D_{\ap}^{\,\beta}\cdot\delta^{\ap\ap}\,.\label{eq:1.15}
\end{align}

For a \textsc{Cartesian} coordinate system the metric tensor components coincide with the \textsc{Kronecker} symbol. Indeed,
\begin{equation*}
  g_{\ip\jp}\ \eqdef\ \i_{\ip}\scal\i_{\jp}\ \equiv\ \delta_{\ip\jp}\,, \qquad
  g^{\ip\jp}\ \eqdef\ \i^{\ip}\scal\i^{\jp}\ \equiv\ \delta^{\ip\jp}\,.
\end{equation*}
Consequently, when we change the coordinates from \textsc{Cartesian} to curvilinear, the metric tensor components are transformed according to the following rule:
\begin{equation*}
  g_{\,ij}\ =\ \D_{\,i}^{\ip}\D_{\,j}^{\jp}\cdot g_{\ip\jp}\,, \qquad
  g^{\,ij}\ =\ \D^{\,i}_{\ip}\,\D^{\,j}_{\jp}\cdot g^{\ip\jp}\,.
\end{equation*}

Let us take an arbitrary vector $\v\ \in\ \R^4\,$. As an element of a vector space it does not depend on the chosen coordinate basis. This object remains the same in any basis\footnote{This statement applies to any other tensor of the first rank.}. However, for simplicity, it is easier to work with vector $\v$ coordinates, which are changing from one basis to another. Let us develop vector $\v$ in three coordinate systems:
\begin{equation*}
  \v\ =\ v_{\ip}\,\i^{\ip}\ =\ \ups_{\,i}\,\e^{\,i}\ =\ \ups^{\,i}\,\e_{\,i}\,.
\end{equation*}
Consequently, $\{v_{\,\ip}\}_{\ip\,=\,0}^{3}$ are \textsc{Cartesian}, $\{\ups_{\,i}\}_{i\,=\,0}^{3}$ are covariant and $\{\ups^{\,i}\}_{i\,=\,0}^{3}$ are contravariant components of the same vector $\v\ \in\ \R^4\,$. As we said, a vector $\v$ as an element of a vector space is invariant \cite{Vekua1978}. However, sometimes we shall say that a vector is covariant (contravariant) by meaning that this vector is given in terms of its covariant (contravariant) components. Using aforementioned relations between various bases vectors we can obtain relations among the covariant and contravariant components solely \cite{Vekua1978}:
\begin{equation}\label{eq:cocontra}
  \ups_{\,i}\ =\ g_{\,ij}\,\ups^{\,j}\,, \qquad
  \ups^{\,i}\ =\ g^{\,ij}\,\ups_{\,j}\,,
\end{equation}
or express covariant (contravariant) components through respective \textsc{Cartesian} coordinates:
\begin{equation*}
  \ups_{\,i}\ =\ \D_{\,i}^{\ip}\,v_{\ip}\,, \qquad
  \ups^{\,i}\ =\ \D^{\,i}_{\ip}\,v^{\ip}\,,
\end{equation*}
and inversely we can express the \textsc{Cartesian} coordinates through the covariant (contravariant) components of the vector $\v\,$:
\begin{equation*}
  v_{\ip}\ =\ \D_{\ip}^{\,i}\,\ups_{\,i}\ \equiv\ v^{\ip}\ =\ \D_{\,i}^{\ip} \ups^{\,i}\,.
\end{equation*}
Two last sets of relations express actually the transformation rules of a general $1-$tensor from a coordinate system to another one. Notice also that
\begin{equation*}
  v_{\Op}\ =\ v^{\Op}\ =\ \od{x^0}{t}\ =\ 1\,, \qquad \mbox{ thus, } \qquad \ups^{\,0}\ \equiv\ 1\,.
\end{equation*}

Let us denote by $\J$ the \textsc{Jacobian} of the transformation \eqref{eq:direct}:
\begin{equation}\label{eq:jacobian}
  \J\ \eqdef\ \det\,\bigl\{\D_{\,i}^{\ip}\bigr\}\,.
\end{equation}
It is not difficult to show that we have also
\begin{equation*}
  \J\ =\ \det\,\bigl\{\D_{\,\alpha}^{\ap}\bigr\}\,, \qquad
  \abs{\J}\ =\ \sqrt{\det\,\bigl\{g_{\,ij}\bigr\}}\ =\ \frac{1}{\sqrt{\det\,\bigl\{g^{\,ij}\bigr\}}}\,.
\end{equation*}
Also from \textsc{Cramer}'s rule we have
\begin{equation*}
  \D_{\ap}^{\,\alpha}\ =\ (-1)^{\alpha - \ap}\;\frac{\det\,\bigl\{\D_{\,\gamma}^{\gp}\bigr\}}{\J}\,, \qquad \gamma\ \neq\ \alpha\,,\quad \gp\ \neq\ \ap\,.
\end{equation*}
For the sequel we shall need to introduce the so-called \textsc{Christoffel} symbols. First we introduce vectors
\begin{equation*}
  \e_{\,ik}\ \equiv\ \e_{\,ki}\ \eqdef\ \pd{\,\e_{\,i}}{q^k}\ \equiv\ \pd{\,\e_{\,k}}{q^i}\,.
\end{equation*}
The expansion coefficients of these vectors in contravariant and covariant bases
\begin{equation*}
  \e_{\,ij}\ =\ \G_{\,ij,\,k}\,\e^{\,k}\ =\ \G_{\,ij}^{\,k}\,\e_{\,k}
\end{equation*}
are called \textsc{Christoffel} symbols of the first and second kind correspondingly:
\begin{equation*}
  \G_{\,ij,\,k}\ \eqdef\ \e_{\,ij}\scal\e_{\,k}\,, \qquad
  \G_{\,ij}^{\,k}\ \eqdef\ \e_{\,ij}\scal\e^{\,k}\,.
\end{equation*}
It is not difficult to see that for a \textsc{Cartesian} coordinate system all \textsc{Christoffel} symbols are identically zero. In the general case \textsc{Christoffel} symbols satisfy the following relations
\begin{equation*}
  \G_{\,ij,\,k}\ \equiv\ \G_{\,ji,\,k}\,, \qquad
  \G_{\,ij}^{\,k}\ \equiv\ \G_{\,ji}^{\,k}\,, \qquad \G_{\,jk}^{\,0}\ \equiv\ 0\,,
\end{equation*}
\begin{equation*}
  \G_{\,kj,\,i}\ +\ \G_{\,ij,\,k}\ =\ \pd{g_{\,ik}}{q^{\,j}}\,, \qquad
  \G_{\,jk}^{\,i}\ =\ \D_{\jp}^{\,i}\cdot\pd{\D_{\,j}^{\jp}}{q^{\,k}}\,,
\end{equation*}
\begin{equation*}
  \G_{\,ik,\,j}\ =\ \frac{1}{2}\;\Bigl[\,\pd{g_{\,ij}}{q^{\,k}}\ +\ \pd{g_{\,kj}}{q^{\,i}}\ -\ \pd{g_{\,ik}}{q^{\,j}}\,\Bigr]\,, \qquad
  \G_{\,ik,\,j}\ =\ g_{\,\ell j}\,\G_{\,ik}^{\,\ell}\,, \qquad
  \G_{\,ik}^{\,\ell}\ =\ g^{\,\ell j}\,\G_{\,ik,\,j}\,.
\end{equation*}
Using the definition of the \textsc{Jacobian} $\J$, it is straightforward to show the following important identity:
\begin{equation}\label{eq:jac}
  \G_{\,ik}^{\,i}\ =\ \frac{1}{\J}\;\pd{\J}{q^{\,k}}\,.
\end{equation}

In curvilinear coordinates the analogue of a partial derivative with respect to a coordinate (\ie an independent variable) is a covariant derivative over curvilinear coordinates which is defined using \textsc{Christoffel} symbols. For example, the covariant derivative $\grad_k$ of a covariant component $\ups_{\,i}$ of vector $\v$ is defined as
\begin{equation*}
  \grad_k\, \ups_{\,i}\ =\ \pd{\,\ups_{\,i}}{q^{\,k}}\ -\ \G_{\,ik}^{\,j}\,\ups_{\,j}\,.
\end{equation*}
Similarly one can define the covariant derivative of a contravariant component $\ups^{\,i}$ of vector $\v\,$:
\begin{equation*}
  \grad_k\, \ups^{\,i}\ =\ \pd{\,\ups^{\,i}}{q^{\,k}}\ +\ \G_{\,kj}^{\,i}\,\ups^{\,j}\,.
\end{equation*}
We remind that in \textsc{Cartesian} coordinates the covariant derivatives coincide with usual partial derivatives since \textsc{Christoffel}'s symbols vanish. For instance, the divergence of a vector field $\v\bigl(q^0,\,q^1,\,q^2,\,q^3\bigr)\ \in\ \R^4$ can be readily obtained:
\begin{equation*}
  \div\v\ =\ \grad_i\,\ups^{\,i}\ =\ \pd{\,\ups^{\,i}}{q^{\,i}}\ +\ \G_{\,ik}^{\,i}\,\ups^{\,k}\,.
\end{equation*}
The last equation can be equivalently rewritten using formula \eqref{eq:jac} as
\begin{equation}\label{eq:divdef}
  \div\v\ =\ \frac{1}{\J}\;\pd{(\J\,\ups^{\,i})}{q^{\,i}}\,.
\end{equation}

In order to recast \textsc{Euler} equations in arbitrary moving frames of reference, we shall have to work with $2-$tensors as well. Similarly to vectors (or $1-$tensors) these objects are independent from the frame of reference. However, when we change the coordinates, tensor components change accordingly. Suppose that we have a $2-$tensor $\T\ =\ \bigl\{T_{\ip\jp}\bigr\}_{\ip\jp\,=\,0}^{\,3}$ and we know its components $T_{\ip\jp}\ \equiv\ T^{\ip\jp}$ in a \textsc{Cartesian} frame of reference. Then, in any other curvilinear coordinates these components can be computed as
\begin{equation}\label{eq:tensor}
  T_{\,ij}\ =\ \D_{\,i}^{\ip}\,\D_{\,j}^{\jp}\,T_{\ip\jp}\,, \qquad
  T^{\,ij}\ =\ \D^{\,i}_{\ip}\,\D^{\,j}_{\jp}\,T^{\ip\jp}\,.
\end{equation}
Similarly, covariant derivatives of a $2-$tensor components are defined as
\begin{align*}
  \grad_{\,k}\,T_{\,ij}\ &=\ \pd{\,T_{\,ij}}{q^{\,k}}\ -\ \G_{\,ik}^{\,\ell}\,T_{\,\ell j}\ -\ \G_{\,jk}^{\,\ell}\,T_{\,i\ell}\,, \\
  \grad_{\,k}\,T^{\,ij}\ &=\ \pd{\,T^{\,ij}}{q^{\,k}}\ +\ \G_{\,k\ell}^{\,i}\,T^{\,\ell j}\ +\ \G_{\,k\ell}^{\,j}\,T^{\,i\ell}\,.
\end{align*}
Now one can show that covariant derivatives of the metric tensor vanish, \ie
\begin{equation}\label{eq:gij}
  \grad_{k}\,g_{\,ij}\ =\ 0\,, \qquad
  \grad_{k}\,g^{\,ij}\ =\ 0\,.
\end{equation}
Finally, we can compute the divergence of a $2-$tensor, which is by definition a $1-$tensor defined as
\begin{equation*}
  \bigl\{\div\T\bigr\}^{\,i}\ =\ \grad_j\,T^{\,ij}\,.
\end{equation*}
Similarly to the divergence of a vector field, the divergence of a $2-$tensor can be also expressed through the \textsc{Jacobian} thanks to the aforementioned definition and formula \eqref{eq:jac} as
\begin{equation}\label{eq:div}
  \bigl\{\div\T\bigr\}^{\,i}\ =\ \frac{1}{\J}\;\pd{\,(\J\,T^{\,ij})}{q^{\,j}}\ +\ \G_{\,jk}^{\,i}\,T^{\,jk}\,.
\end{equation}

\paragraph*{Flow vorticity.} Using the \textsc{Levi}-\textsc{Civita} tensor $\varepsilon^{\,\alpha\,\beta\,\gamma}\,$, which is defined as \cite{Sedov1997}:
\begin{equation*}
  \varepsilon^{\,\alpha\,\beta\,\gamma}\ =\ \begin{dcases}
    \ \frac{1}{\J}\,, & \sign\begin{pmatrix} 1 & 2 & 3 \\ \alpha & \beta & \gamma \end{pmatrix}\ =\ 1\,, \\
    \ -\frac{1}{\J}\,, & \sign\begin{pmatrix} 1 & 2 & 3 \\ \alpha & \beta & \gamma \end{pmatrix}\ =\ -1\,, \\
    \ 0\,, & (\alpha\ =\ \beta)\ \vee\ (\alpha\ =\ \gamma)\ \vee\ (\beta\ =\ \gamma)\,,
  \end{dcases}
\end{equation*}
where $\sign(\cdot)$ is the signature of a permutation. In other words the sign of \textsc{Levi}-\textsc{Civita} tensor components change the sign depending if the permutation $(\alpha\ \beta\ \gamma)$ is odd or even. Using this tensor, we can define the \emph{rotor} of a vector $\v$ as:
\begin{equation*}
  \bomega\ =\ \rot\v\ =\ \omega^{\,\gamma}\,\e_{\,\gamma}\,, \qquad
  \omega^{\,\gamma}\ \eqdef\ \varepsilon^{\,\alpha\,\beta\,\gamma}\cdot\grad_{\,\alpha}\,\ups_{\,\beta}\,.
\end{equation*}
We can write explicitly the contravariant components of vector $\bomega\,$:
\begin{equation}\label{eq:vort}
  \omega^{\,1}\ =\ \frac{1}{\J}\;\Bigl(\pd{\ups_{\,3}}{q^{\,2}}\ -\ \pd{\ups_{\,2}}{q^{\,3}}\Bigr)\,, \qquad
  \omega^{\,2}\ =\ \frac{1}{\J}\;\Bigl(\pd{\ups_{\,1}}{q^{\,3}}\ -\ \pd{\ups_{\,3}}{q^{\,1}}\Bigr)\,, \qquad
  \omega^{\,3}\ =\ \frac{1}{\J}\;\Bigl(\pd{\ups_{\,2}}{q^{\,1}}\ -\ \pd{\ups_{\,1}}{q^{\,2}}\Bigr)\,.
\end{equation}

These technicalities are used in our paper in order to reformulate the full \textsc{Euler} equations in an arbitrary moving curvilinear frame of reference. As a practical application we employ these techniques to the globally spherical geometry due to obvious applications in Geophysical Fluid Dynamics on the planetary scale.


\section{Acronyms}

In the text above the reader could encounter the following acronyms:

\begin{description}
  \item[NSW] Nonlinear Shallow Water
  \item[PDE] Partial Differential Equation
  \item[TVD] Total Variation Diminishing
  \item[DART] Deep-ocean Assessment and Reporting of Tsunamis
  \item[MOST] Method of Splitting Tsunami
  \item[NSWE] Nonlinear Shallow Water Equations
\end{description}


\bigskip\bigskip
\addcontentsline{toc}{section}{References}
\bibliographystyle{abbrv}

\bigskip\bigskip

\end{document}